\documentclass[a4paper,fleqn,usenatbib]{mnras}

\usepackage{bm}

\usepackage[T1]{fontenc}
\usepackage{ae,aecompl}

\usepackage{graphicx}	
\usepackage{amsmath}	
\usepackage{amssymb}	

\newcommand{\eqb}{\begin{eqnarray}}
\newcommand{\eqe}{\end{eqnarray}}
\newcommand{\eqbn}{\begin{eqnarray*}}
\newcommand{\eqen}{\end{eqnarray*}}
\newcommand{\eqnb}{\begin{equation*}}
\newcommand{\eqne}{\end{equation*}}

\newcommand\comment[1]{{#1}}

\title[CR current-driven instabilities near sources]{Cosmic-ray current-driven instabilities -- revisiting environmental conditions}

\author[Reville, Giacinti \& Scott]{
Brian Reville$^{1}$\thanks{E-mail: brian.reville@mpi-hd.mpg.de }
Gwenael Giacinti$^{1}$ and Robyn Scott$^{2}$
\\
$^{1}$Max-Planck-Insitut f\"ur Kernphysik, Saupfercheckweg 1, Heidelberg 69117, Germany\\
$^{2}$ Centre for Plasma Physics, Queen's University Belfast, University Road, Belfast BT7 1NN, Northern Ireland
}

\date{Accepted 2021 January. Received 2021 January; in original form 2020 November }

\pubyear{2020}

\begin{document}
\label{firstpage}
\pagerange{\pageref{firstpage}--\pageref{lastpage}}
\maketitle

\begin{abstract}
The growth of magneto-hydrodynamic fluctuations relevant to cosmic ray confinement in and near their sources, and the effects of local plasma conditions is revisited. We consider cases where cosmic rays penetrate a medium which may contain a fraction of neutral particles, and explore the possible effects of high-order cosmic-ray  anisotropies. An algorithm for calculating the dispersion relation for arbitrary distributions, and anisotropies is presented, and a general solution for power-law cosmic-ray distributions is provided. Implications for the resulting instabilities near to strong Galactic cosmic-ray sources are discussed. We argue that cosmic-ray streaming in weakly ionised plasmas eliminates the need for the existence of an evanescent band in the dispersion relation, a conclusion which may be confirmed by gamma-ray observations. The necessity for additional multi-scale numerical simulations is highlighted, as understanding the non-linear behaviour is crucial. 
\end{abstract}

\begin{keywords}
 plasmas -- instabilities -- (ISM:) cosmic rays
\end{keywords}



\section{Introduction}

Young supernova remnants (SNR) are sources of high-energy particles, revealed by their non-thermal x-ray and $\gamma$-ray emission \cite[e.g.][]{Berezhko,HintonHof}. If they are also sources of cosmic rays (CRs), as is likely the case, then the escape and transport in their local environment should hint at the processes determining confinement, and offer a potential probe of the time-history and conditions near the source. Such information is required to understand their contribution to the total Galactic CR population. 

Studies on CR escape from sources, dating back to the early works of \citet{KulsrudCesarsky71}, \citet{Skilling71} and others, implicitly assume that CR transport, at least at low energies, is dominated by scattering on resonantly-excited Alfv\'en waves.  Close to efficient CR sources, this is however unlikely to be the case. For example, on scales relevant to the energetic particles emitting the $>$TeV $\gamma$-rays now detected from several young SNRs, estimates for the growth-time of resonantly excited Alfv\'en modes can exceed the lifetime of the SNR in question. The present understanding is that the non-linear development of a much faster non-resonant short-wavelength instability provides the required scattering and confinement \citep{Bell04,Bell13}. 

The concept of magnetic field amplification, and self-confinement underpins all current approaches to model acceleration of CRs to the highest energies in SNRs. It will likewise be a requirement of any conceivable alternative source. The model put forward by \cite{ZirakashviliPtuskin} and \cite{Bell13} establishes an upper-limit to the maximum achievable CR energy; directly related to the ability of escaping high-energy particles to amplify the magnetic field (non-resonantly) to a level that inhibits further escape. In this model, the system is continuously leaking a fraction of particles at the highest energies, their mean free path growing with distance from the shock. The picture bears similarities with the free escape boundary approach commonly used as a closure approximation in non-linear steady state models \cite[e.g.][]{EllisonEichler, Reville09}. 

\comment{This slightly artificial picture is conceptually simple in that it conveniently separates the energetic particles into two populations; one being transported in a highly developed and likely complex field topology close to the shock, and another streaming freely into an ambient medium. 
The former regime has been the focus of numerous numerical investigations \cite[e.g.][]{Bell04,Zirakashvili,Reville08,Gargate,Rogachevskii,Reville13,Baietal15, Marret} although the inherent multi-dimensional, non-linear and multi-scale nature of the problem still poses serious challenges. 
For the latter scenario, capturing the transport and self-regulated scattering in the immediate surrounding medium is an equally challenging problem, as an accurate model demands connecting the global structure of the accelerator with the local large scale turbulent magnetic field, neither of which are completely understood. Ultimately a unified theory that self-consistently captures the complete global picture bridging the two regimes is desired, but a clear understanding of the underlying assumptions is an essential first step.}

Using test particle simulations in an isotropic Kolmogorov turbulent field, \cite{Giacinti} demonstrated that in the process of diffusing away from a source, before CRs have diffused over several correlation lengths $L_c$ of the large scale field, the transport is highly filamentary. The filaments develop due to the anisotropic nature of particle diffusion in magnetised plasmas \cite[e.g.][]{Isenberg}. \cite{Giacalone} has shown that similar filaments may also develop on smaller scales close to the shock, for a given field realisation. Efforts to incorporate this phenomenon into the standard picture have been applied, although for practical purposes in a reduced one-dimensional picture \citep{Ptuskin,Malkovetal} while other have included the role of neutral particles in damping/stabilising the self-excited modes \citep{Nava16,DAngelo,Brahimi20}. These works have exclusively worked within the confines of the standard quasi-linear theory framework discussed previously. \cite{Bell13} on the other hand, made a first attempt to apply advances in our understanding of non-resonant excitation of magnetic fields into a more self-consistent picture of escape from SNRs. This same approach has also been applied by \citet{Blasietal} to the escape of CRs from galaxies.

In this work, we revisit the linear analysis of unstable modes, developing a more generic framework for treating the linear perturbations, with the aim of providing fresh insight into the problem, while also laying the groundwork for future numerical investigations. The outline of the paper is as follows. In the next section we describe the general framework, including the key assumptions. Section 3 details application of the new results to ionised plasmas, using new results from Cas A to motivate the parameter choices. In section 4, the combined effects of neutrals and cosmic-rays are investigated. Due to the lengthy nature of the calculation, the derivation of the cosmic-ray response is left to the appendix.

\section{Cosmic-ray anisotropy and generalised MHD instability}

We consider the generic case of transport in the neighbourhood of a strong CR source. A magneto-hydrodynamic description for the background plasma is adopted, which for simplicity we take to be composed of electrons and protons only. Combining the single fluid equations, while accounting for the presence of cosmic-rays and neutral hydrogen, the momentum equation for the charge carrying thermal fluid reads \citep{Bell04,Reville07}
\eqb
\label{momEqn}
&&\rho_{\rm i} \frac{\mathrm{d} \boldsymbol{u}_{\rm i}}{\mathrm{d} t} =-\nabla P_{\rm th}-\frac{1}{4 \pi} \boldsymbol{B} \times(\nabla \times \boldsymbol{B})-\rho_{\rm i}\nu_{\mathrm{in}}\left(\boldsymbol{u}_{\mathrm{i}}-\boldsymbol{u}_{\mathrm{n}}\right)
   \enspace \enspace \enspace \nonumber \\
 & &   \enspace \enspace \enspace    \enspace \enspace \enspace   \enspace \enspace \enspace    \enspace \enspace \enspace -
\frac{1}{c}(\boldsymbol{j}_{\mathrm{cr}}-n_{\mathrm{cr}} q_{\rm cr} \boldsymbol{u}_{\rm i}) \times \boldsymbol{B} .
\eqe
The last term  ($n_{\mathrm{cr}} q_{\rm cr} \boldsymbol{u}_{\rm i}\times  \boldsymbol{B}$) arises due to the fact that the return current is drawn relative to the background fluid, and it is this term that ultimately determines the neutral streaming velocity in the fully ionised case. $\nu_{\rm in}$ is the momentum exchange rate due to elastic charge-exchange collisions with neutral hydrogen atoms. \comment{A derivation of this equation is provided in Appendix A, where previous kinetic approaches are also reviewed.}

Adopting ideal MHD, Faraday's equation reads
\eqb
\frac{\partial \boldsymbol{B}}{\partial t}=\bm{\nabla}  \times( \boldsymbol{u}_{\rm i} \times \boldsymbol{B})~.
\eqe
The above equations must be solved together with momentum conservation for the neutral fluid 
\eqb
 \rho_n\frac{\mathrm{d} \boldsymbol{u}_{\rm n}}{\mathrm{d} t }= -\nabla P_{\rm n} -{\rho}_{\rm i}\nu_{\mathrm{in}}\left(\boldsymbol{u}_{\mathrm{n}}-\boldsymbol{u}_{\mathrm{i}}\right) ~.
\eqe
where from momentum conservation, we have $\rho_{\rm n}\nu_{\rm ni} = {\rho}_{\rm i}\nu_{\mathrm{in}}$.
The following expression for the ion-neutral collision frequency, valid in the range $10^2 {\rm K} \ll T < 10^6 {\rm K}$ \citep{KulsrudCesarsky71} is used
\eqb
\nu_{\rm in} = 8.9\times 10^{-9} n_{\rm n} {T_{\rm i}}^{0.4} ~{\rm s}^{-1} ~,
\eqe 
where $T_{\rm i}$ is the ion temperature in eV. For lower temperatures $T \approx 10^2$ K, we use the results of \cite{Osterbrock} which, since we neglect Helium contributions throughout this work, is simply  
\eqb
\nu_{\rm in} = 2.3\times 10^{-9} {n}_{\rm n} ~{\rm s}^{-1} ~.
\eqe
We take this value to hold at lower temperatures, since the same physical arguments presented in \cite{Osterbrock} apply. Note however, for such small temperature the weakly collisional regime may come in to play, which requires a different analysis. \cite{Bell20} have recently demonstrated that a related instability, the magneto-collisional instability, occurs in this regime.

It is also known that the environments of SNRs have significant amounts of dust, as do many parts of the ISM. A framework for adding this additional charged species to the system of equations and a detailed linear analysis has been performed by \cite{squire}. We will not consider this additional complication here.

We proceed with a standard linear analysis, looking at circularly polarised modes propagating parallel to a mean background field {with average CR current along the mean magnetic field direction: $\bm{j}_0 \| \bm{B}_0$}. Following \cite{Bell04}, we make the substitution $j_\bot/j_0 = \sigma  B_\bot/B_0$
i.e. $\sigma$ describes the response of the CR current to small transverse fluctuations in the magnetic field. 
Working in the local fluid rest frame ($\langle \bm{u}_{\rm i}\rangle = 0$), and excluding any possible drift between neutrals and the ionised fluid to lowest order, the general dispersion relation is
\begin{align}
\label{DispMaster}
&\omega^2 \left[1+\frac{{\rm i} \nu_{\rm in}}{\omega+{\rm i}\chi  \nu_{\rm in}}
\right]=\nonumber\\ 
& ~~~~~~~
k^2 v_{\rm A}^2 - \epsilon\frac{kB_0 j_0}{\rho_{\rm i} c}\left[1-\sigma_1 - \frac{\omega}{k v_{\rm cr}}(1-\sigma_2)\right] ~,
\end{align}
where $v_{\rm A}= B_0/\sqrt{4 \pi \rho_i}$ is the Alfv\'en velocity. The cosmic-ray drift velocity $v_{\rm cr} = j_0/n_{\rm cr} q_{\rm cr} $ is introduced here, as well as an ionisation ratio $\chi={\rho_{\rm i}}/{\rho_{\rm n}}=\nu_{\rm ni}/\nu_{\rm in}$.\footnote{
As opposed to the ionisation \emph{fraction} $X \equiv\rho_{\rm i}/\rho_{\rm tot} = \chi/(1+\chi)$} 
For convenience, we have also separated out the $\omega$ dependent terms as $\sigma(\omega, k) = \sigma_1(k) -(\omega/v_{cr} k) \sigma_2(k)$ (see Appendix B). 

A generalised expression for $\sigma$ is given in the Appendix where an expansion of the cosmic-ray anisotropy to arbitrary order in Legendre polynomials is presented. Other basis functions (e.g. Chebyshev) are equally straightforward to derive, however, here we aim to develop a framework to compare to future simulations which will build on the work of \citet{Reville13} who used the related spherical harmonic functions as a basis. The penultimate expression for $\sigma$, which involves an integral over the magnitude of momentum, is given in a form that allows for straight-forward numerical calculation which can be applied to an arbitrary distribution. In the present work, we focus on a fixed power-law distribution at all orders of the expansion $f_\ell \propto p^{-s}\Theta(p;p_1,p_2)$, where $\Theta(x;a,b)$ is a top hat function, equal to unity for $a<x<b$ and zero otherwise. In this situation, the integrals, although cumbersome, can still be expressed in terms of standard functions, and can be shown to reduce to previous results in the appropriate limits.  e.g. first order expansions with $s=4$ were originally presented  in \cite{Bell04}, while thermal/composition effects, and alternative power-laws have been explored in \cite{Reville07}. We revisit and extend these results in the following.

Higher order anisotropies (to second order in Legendre polynomials) have been presented in \cite{BykovMNRAS11}, although there the emphasis was on the firehose instability. Using the asymptotic $kr_{\rm g,2}\ll1$ values given in the appendix, where $r_{\rm g,2}=r_{\rm g}(p_2)$ is the gyro-radius of the maximum energy cosmic-rays, independent of $s$ the usual firehose expression is found
\eqb
\omega^2 = k^2 v_{\rm A}^2\left( 1 -\frac{P_{\|}-P_\bot}{B_0^2/4\pi}\right)~,
\eqe
where $P_{\|,\bot}$ are the parallel,perpendicular components (w.r.t. the guide field) of the cosmic-ray pressure tensor  \citep{Bykovetal11,Scottetal}. If the pressure anisotropy $\Delta P_{\rm cr} = P_\|-P_\bot$ is large enough the system is firehose unstable. We will see that for what concerns the acceleration of the highest energy particles, this regime is generally too slow to be of any influence, at least for the cases we consider.

We may also compare our result to that of the resonant ion-cyclotron instability, valid in the weak damping approximation. As we will show below, this case applies in the limit ${v_{\rm cr}}/{c} \ll U_{B}/U_{\rm cr}$, where $U_B$ and $U_{\rm cr}$ are the energy densities of the magnetic field and CRs respectively. In this limit the first term on the rhs of equation \ref{DispMaster} dominates and the real part of the dispersion relation corresponds to left or right propagating  Alfv\'en waves ${\rm Re}({\omega})= \pm k v_{\rm A}$.
Using the asymptotic expressions given in equations (\ref{sig1-res}) and (\ref{sig2-res}) of the Appendix,
we recover the familiar linear growth rate expression for the resonant streaming instability
\begin{align}
&{ \rm Im}({\omega}) = \pm {\Omega_0} \left.\frac{\pi}{4} \frac{s-3}{s-2}
\frac{n_{\rm cr}\left(>p\right)}{n_i }\left(\frac{3}{s}\frac{v_{\rm cr}}{v_{\rm A}}\mp  1\right)\right|_{p=eB/ck} ~,\nonumber
\end{align}
where $\Omega_0= eB_0/m_p c$ is the non-relativistic gyro-frequency, and $n_{\rm cr}(>p) = 4 \pi \int_p^\infty f p^2dp$.
Note that only waves with $\omega/k$ and $v_{\rm cr}$ of the same sign have positive growth. We see here also the neutral streaming speed has the correct form, \cite[e.g.][]{MelroseWentzel70}, namely that steady streaming is proportional to the Alfv\'en velocity. If the self-generated waves reduce the streaming velocity to the threshold condition $v_{\rm cr} = s v_{\rm A}/3$, the next order term in the anisotropy can be included. Using the results in the appendix B5, the pressure anisotropy instability recently investigated by \cite{Zweibel20} is recovered. The instabilities are not mutually exclusive, but rather limiting forms of the resonantly driven instability for vanishing first and second order anisotropies respectively. 

Interest in cosmic-ray streaming has been renewed in recent years, where 
{ significant efforts have been made to implement the process in large galaxy-scale 
simulations} \citep{JiangOh, ThomasPfrommer, Hopkins}. In this regard, it is necessary to highlight the range of validity of the streaming instability result above. The previously mentioned limit on applicability  of the weak-damping approximation corresponds to a cosmic-ray energy density of $U_{\rm cr} \ll 10 ~  (v_{\rm cr}/v_{\rm A})^{-1} n_{gas}^{1/2} B_{\rm \mu G}$~keV~cm$^{-3}$. This limit has a simple physical interpretation, being equivalent to the condition that the $\bm{j}_{cr} \times \bm{B}$ force on the background fluid does not dominate over the magnetic tension (a necessary condition for Alfv\'en wave propagation) on any scale in the resonant range. The above limit is not always guaranteed to be the case, particularly in the neighbourhoods of strong CR sources, where both $U_{\rm cr}$ and $v_{\rm cr}/v_{\rm A}$ are enhanced relative to Galactic averages.
When this condition is violated, the short-wavelength non-resonant branch first investigated by \cite{Bell04} dominates. However, as we show below, the resonant modes also become increasingly sensitive to the nature and details of the anisotropy.

With the goal of generalising the prescription for parallel propagating modes in magnetised plasmas containing a population of drifting CRs in mind, it will prove convenient to introduce the dimensionless variables $(\hat{\omega}, \hat{\nu}) = (\omega, \nu_{\rm in}) \times  r_{\rm g1}/v_{\rm A}$ and $\hat{k} = k r_{\rm g1}$, where $r_{\rm g1}$ is the gyro-radius of the lowest energy cosmic rays in the background mean field. We 
keep the treatment general, such that $v_{\rm cr}$ can take any value, but will specialise later to the case of shock precursors.

In dimensionless form, equation (\ref{DispMaster}) reads
\eqb
\label{DispDimensionless}
\hat{\omega}^2 \left[1+\frac{{\rm i} \hat{\nu}}{\hat{\omega}+{\rm i}\chi \hat{ \nu}}
\right]= \hat{k}^2  - \epsilon \zeta_1 \hat{k} \left[1-\sigma_1 - \frac{\hat{\omega}}{\hat{k}} \frac{v_{\rm A}}{v_{\rm cr}}(1-\sigma_2)\right]~,
\eqe
where 
we have introduced the dimensionless quantity
\begin{align}
\zeta_1 &= \frac{r_{\rm g1} B_\| j_\|}{\rho_{\rm i} v_{\rm A}^2 c} =
\frac{4 \pi n_{\rm cr} p_1 v_{\rm cr}}{B_\|^2} \nonumber \\
&=\frac{1}{2}\frac{(s-4)}{(s-3)}
\frac{1-\left(\frac{p_1}{p_2} \right)^{s-3}}{1-\left(\frac{p_1}{p_2} \right)^{s-4}}\left(\frac{U_{cr}}{U_{\rm B}}\right) \frac{v_{\rm cr}}{c} ~.
\label{zetaA}
\end{align}
In this notation, the condition for Bell-instability to operate is $\zeta_1>1$. If this condition is not satisfied, the magnetic tension dominates at $k r_{\rm g,1} = 1$, and the non-resonant branch is effectively stabilised. The weak-damping result given above generally applies in this case, provided the anisotropy is  
small. If the anisotropy is large, specifically if $\Delta P_{\rm cr}/ P_{\rm cr} > v_{\rm cr}/{c}$, the leading order terms are comparable. The effect of increased anisotropy will be shown in the next section.
The second form for $\zeta_1$ given in (\ref{zetaA}) above is close\footnote{
For the  $s=4$ case, note $\lim_{n\rightarrow 0} \frac{x^{n}-1}{n} = \ln x$ } to that used in \cite{Bell04} although the assumption that $v_{\rm  cr} = u_{\rm sh}$ is no longer implicit. Note that setting $U_B$ to the total magnetic energy (turbulent + mean field) in this expression, the saturation condition given in \cite{Bell04} corresponds to $\zeta_1=1$.

An alternative approach introduced by \cite{Bell13}, which is more applicable to CR escape upstream of a shock, is to define an escaping flux $j_{\|} \bar{p}_{\rm esc} c =  \eta_{\rm esc} e \rho_{\rm tot} u_{\rm sh}^3$, where $\eta_{\rm esc}$ represents the fraction of the total mass energy density processes by the shock being carried away by escaping particles, and $\bar{p}_{\rm esc}$ the average momentum of these escaping particles. Setting this to $p_1$, we find 
\eqb
\label{zeta3}
\zeta_1 = \eta_{\rm esc} M_{\rm A,n}^2 \frac{u_{\rm sh}}{c} ,
\eqe
where $M_{\rm A,n} = u_{\rm sh}/v_{\rm A,n}$ is the Mach number of the shock. Here we introduce the reduced Alfv\'en velocity $v_{\rm A,n} = B/\sqrt{4 \pi (\rho_{\rm i}+\rho_{\rm n})}$ in anticipation of shocks that encounter an incompletely ionised medium in the far upstream. This reduces to the expression in \cite{Bell13} in the limit $\rho_n = 0$. 

We explore the new generalised dispersion relation below for a selection of specific relevant cases.

\begin{figure*}
	\begin{center}
		\includegraphics[width=\columnwidth]{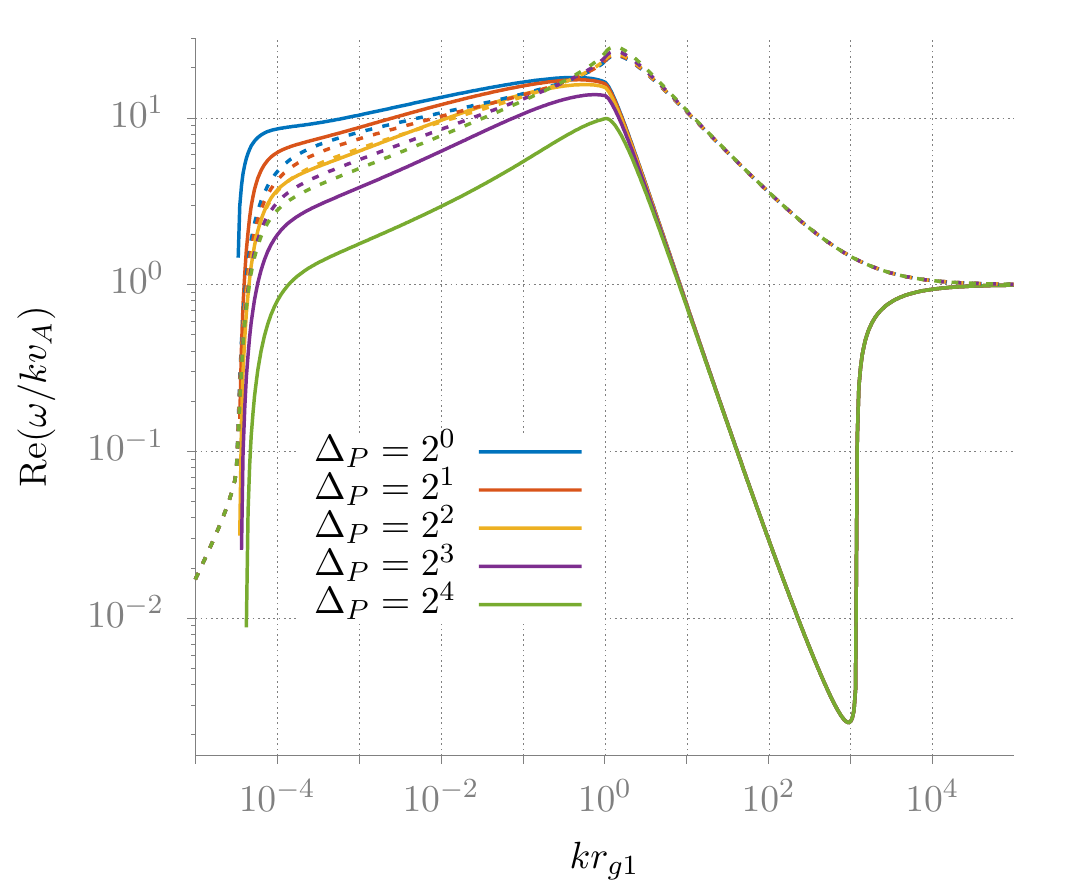} 
		\includegraphics[width=\columnwidth]{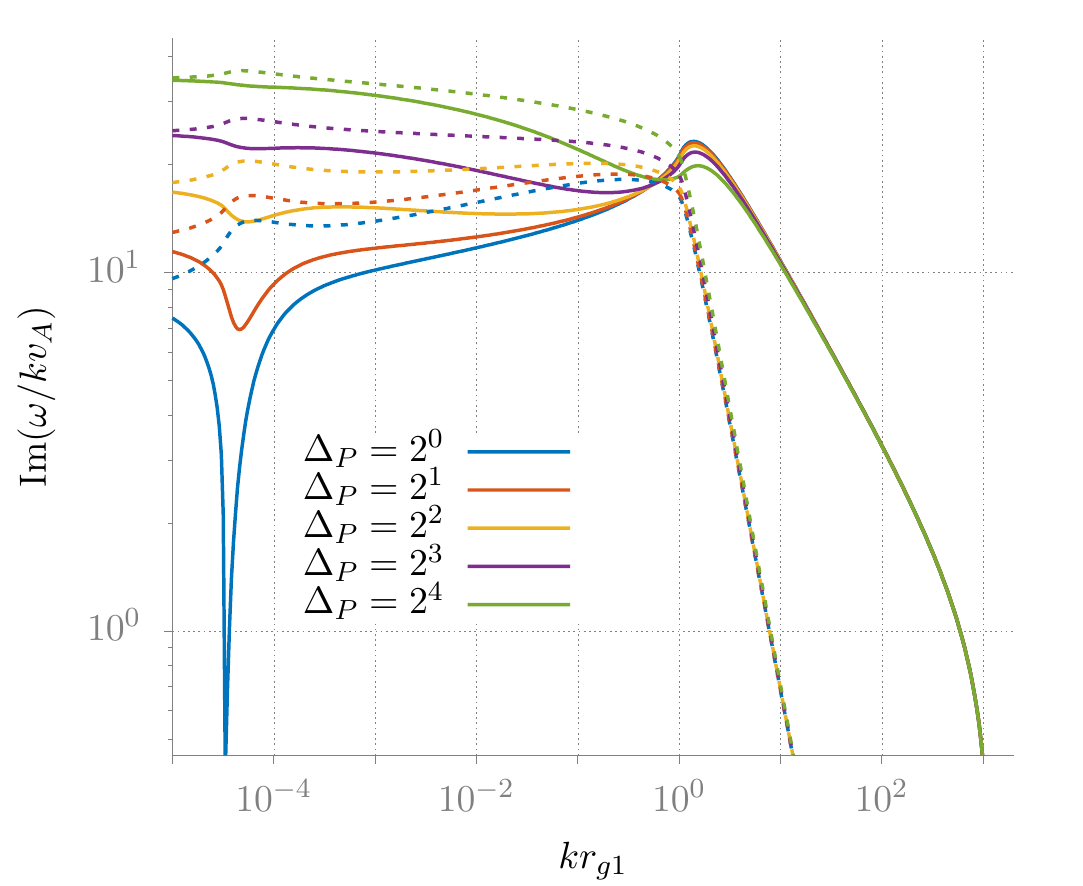}
	\end{center}
	
	\caption{Real (left) and imaginary (right) components of the phase speeds for waves in a fully ionised background, normalised to the ambient Alfv\'en velocity, as a function of pressure anisotropy $\Delta_P = (\Delta P_{\rm cr}/P_{\rm cr})/(2 (u_{\rm sh}/c)^2)$ (i.e. the fractional pressure anisotropy relative to the diffusion approximation result). Dashed curves are for modes rotating in the same sense as the driving cosmic-rays, while solid lines in the opposite sense. We use values expected for Cas A:  $v_{\rm cr}=u_{\rm sh} = 5,000~{\rm km~s}^{-1}$, $\rho_{\rm i} = 2.34\times 10^{-24}$ g  cm$^{-3}$ and $B_0 = 3~\mu$G. We take $p_{2}/p_{1} = 10^{4.5}$, such that for $E_{\rm min}=1$GeV, $E_{\rm max}\approx 30$ TeV. We have taken a CR efficiency of $P_{\rm cr}/\rho_{\rm i} u_{\rm sh}^2 = 0.2$ (see text). Note the cut-off in the real part of the frequency below $k r_{\rm g2} = 10^{-4.5}$ due to onset of the firehose instability.}
	\label{fig:1}
\end{figure*}

\section{Strongly ionised plasmas - $\chi \rightarrow \infty$}
\label{sec:3}

Observations of several SNRs in H$\alpha$ provide unambiguous evidence that SNR shocks interact with partially ionised plasmas. As previously pointed out by \cite{Reville13}, the strong heating in the non-linear phase of CR-driven magnetic field amplification is likely to ionise most material before it reaches the shock, subject to the obvious condition that the driving is strong enough to reach this stage of non-linear development. The apparent anti-correlation between synchrotron X-ray rims, a tracer of strong magnetic field amplification, and H$\alpha$ emission in some cases \citep{Winkler} indicate that this might be the case. Hence, the completely ionised scenario likely applies to the developed precursors of strong SNR shocks. 
Here we apply the new generalised expressions, motivated by recent observational results which better constrain the cosmic-ray parameters.

\subsubsection*{Cas A:}

Recent observations by \cite{CasAMAGIC} and in particular the combined Fermi-VERITAS analysis of \cite{CasAVeritas} provide robust constraints on the CR parameters in Cas A. While a full numerical model that captures self-consistent acceleration and field growth is desirable, here we simply aim to use the revised dispersion relation to explore the various modes that mediate the scattering. We focus on the pure hadronic model of \citet{CasAVeritas}, which indicates a total CR energy $E_{\rm CR} \approx 1.7 \times 10^{50}$ erg, with best fit above the pion bump exhibiting a power-law with exponent $s=4.17$ and exponential cut-off at $pc=17$ TeV. Adopting for the total available energy of the SNR a value of $E_{\rm SNR} = 2-5 \times 10^{51}$~erg \citep{ChevalierOishi}, the hadronic model implies a total (i.e. intergrated over the SNR lifetime) CR acceleration efficiency $\approx 5\%$. 

To determine the dispersion of the linear modes, we require two additional pieces of information. The cosmic-ray energy density in the precursor, and limits on the higher-order CR anisotropy. For the latter, we will restrict ourselves to second order, i.e. pressure anisotropy.
For a shock radius of $2.5$~pc, adopting the total CR energy in Cas A given by \citet{CasAVeritas}, the average energy density is $\left\langle U_{\rm cr}\right\rangle \approx 60$ keV cm$^{-3}$. This is to be compared with the mean energy density of the shock processed material $\bar{U}_{\rm sh} \equiv \frac{1}{2} \rho_{\rm i} u_{\rm sh}^2 \approx 180$ keV cm$^{-3}$. In the latter estimate, we have adopted the same gas density upstream of the shock $\rho_{\rm i} = 2.34\times 10^{-24}$ g  cm$^{-3}$  as used in \citet{CasAVeritas}. In terms of hydrodynamic feedback, this gives $\eta \equiv P_{\rm cr}/\rho_0 u_{\rm sh}^2 \approx 0.2$. Comparable hydrodynamic efficiencies for other young remnants have been claimed previously in the literature, although here we will adopt it as an upper limit, as it can not yet be ruled out that  most of the energy resides deep in the SNR interior.

\begin{figure*}
	\begin{center}
		\includegraphics[width=\columnwidth]{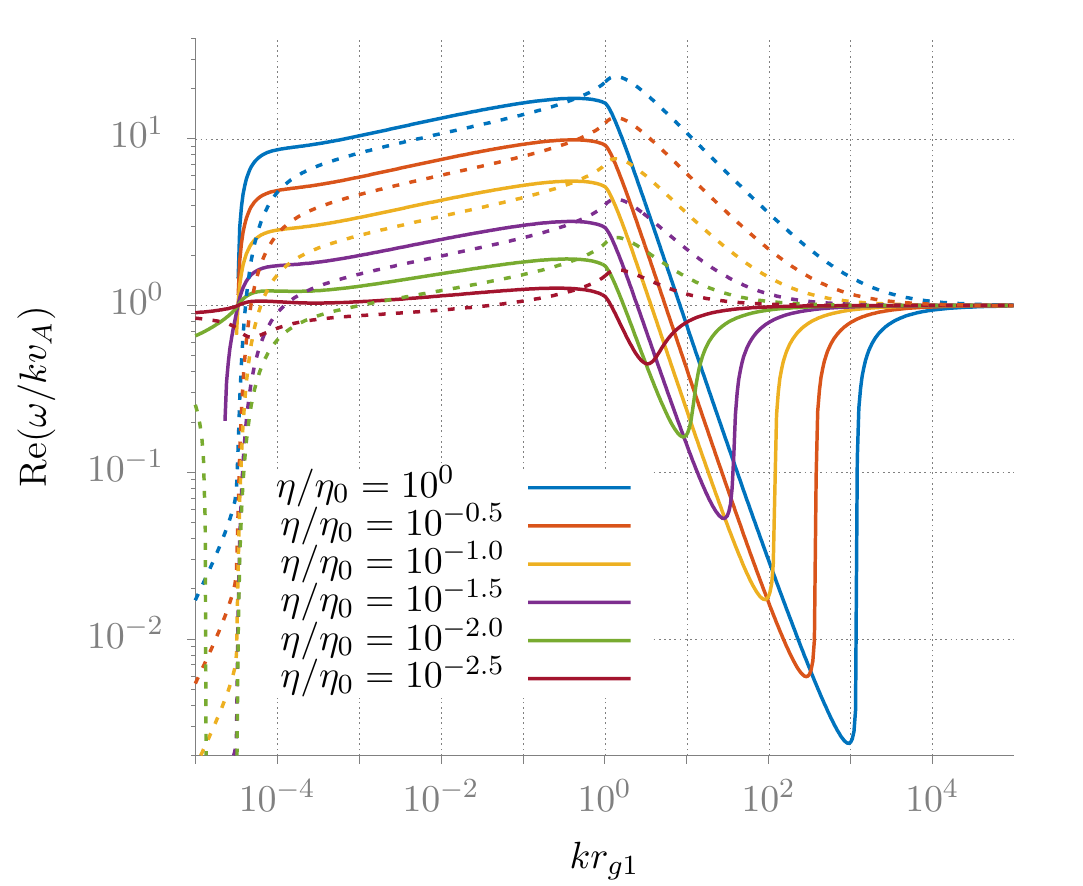}
		\includegraphics[width=\columnwidth]{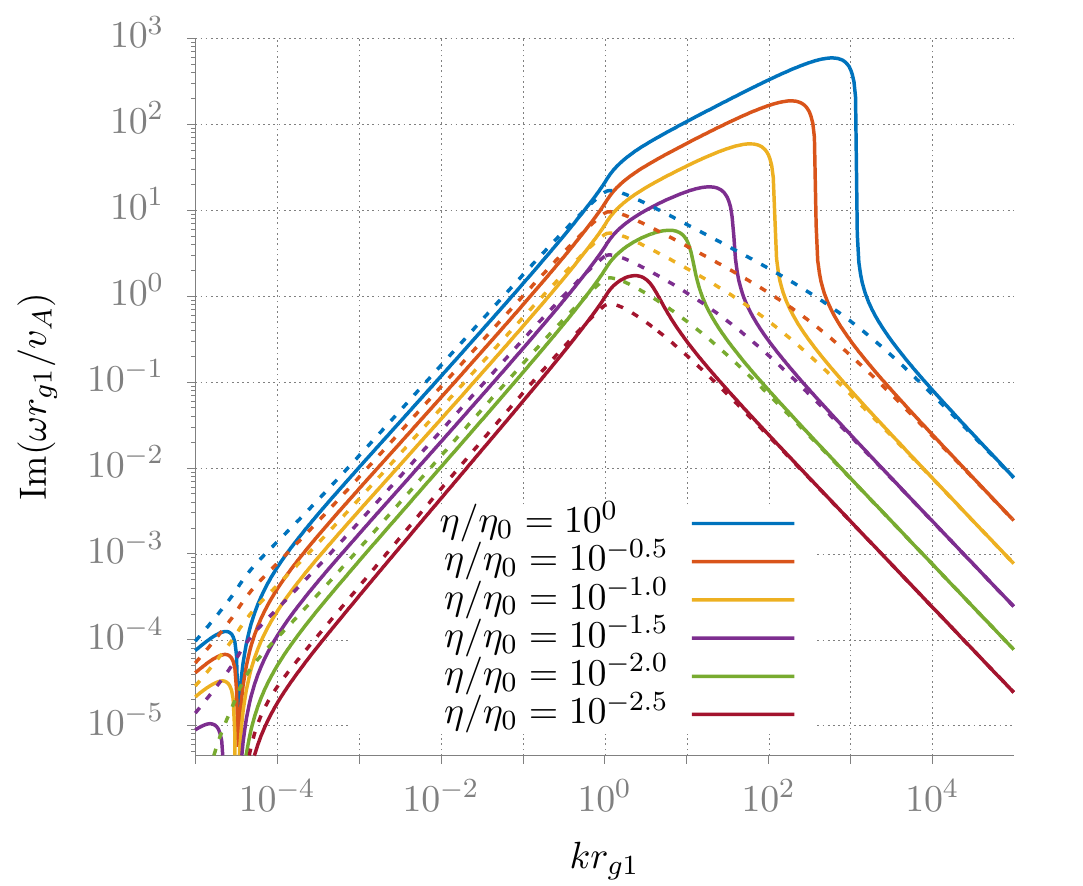}
	\end{center}
	
	\caption{Real component of the phase speeds (left) and growth rates (right) in a fully ionised background and growth rate, as a function of cosmic-ray acceleration efficiency $\eta=P_{\rm cr}/\rho_{\rm i} u_{\rm sh}^2$. Pressure anisotropy is set to the diffusion approximation $\Delta_P = 2 (u_{\rm sh}/c)^2$, and $\eta_0=0.2$ (see text). All other parameters are as in Fig.\ref{fig:1}. In the lower plot, for the conditions given, the relevant time scale for growth is $r_{\rm g1}/v_{\rm A}\approx 10^6 (p_1/m_{\rm p}c)$~s. We recall that the maximum growth rate is predicted to be Im$(\omega r_{\rm g1}/v_{\rm A} = \zeta_1/2)$, where in these plots $\zeta_1 \approx 1200 \eta/\eta_0$}
	\label{fig:2}
\end{figure*}

With this picture in mind, we can repeat the estimates of \cite{Bell13}, assuming the non-resonant mode dominates, and that confinement demands a minimum of $N\approx5$ growth times ($t_{\rm NR} = 2r_{\rm g1}/v_{\rm A}\zeta_1$). In a quasi-steady state, we expect the escaping flux to match the average accelerating flux close to the cut-off
\citep{Drury99},
\begin{align}
j_{\rm acc}(p_{\rm max}) = 
\int_{p_{\rm max}-\Delta p}^{p_{\rm max}} d^3 p~ f(\bm{p}) \bm {v}\cdot\bm{n} 
\approx  \frac{4\pi \Delta u}{3}p_{\rm max}^3 f_0(p_{\rm max})~,
\end{align}
where $\bm{v}\cdot\bm{n}$ is the projection of the particle velocity vector onto the shock normal, $\Delta p$ the average momentum change per cycle, and $\Delta u$ the velocity jump across the shock. Assuming the  cosmic-rays populate a power-law distribution, extending without break to low energies, 
\begin{align}
j_{\rm acc}(p_{\rm max}) = \frac{P_{\rm cr}}{\rho_i u_{\rm sh}^2} n_i \Delta u 
\left(\frac{u_{\rm sh}}{c}\right)^2 \frac{(p_{\rm max}/m_pc)^{3-s}}{F(p_0, p_{\rm max})~~}
\end{align}
with
\begin{align*}
F(p_0, p_{\rm max}) = \int_{p_0/m_pc}^{p_{\rm max}/m_pc}\frac{u^{4-s}}{\sqrt{1+u^2}} ~du~.
\end{align*}
This integral can be expressed as a combination of Hypergeometric functions, but it suffices to know that it is of order unity for any realistic choice of parameters. 
Using this value for the escaping flux of CRs upstream, the confinement condition $t_{\rm NR} < t_{\rm SNR}/N$ thus becomes
\begin{align}
\left(\frac{p_{\rm max}}{m_{\rm p}c}\right)^{s-3} = \frac{3}{4}\frac{P_{\rm cr}}{\rho_i u_{\rm sh}^2} \left(\frac{u_{\rm sh}}{c}\right)^3 \frac{\omega_{\rm pi} t_{\rm SNR}}{N F }~,
\end{align}
where $\omega_{\rm pi}$ is the background ion plasma frequency upstream of the shock. For $\eta=0.2$, 
and $s=4.17$, this gives $E_{\rm max} \approx 50$ TeV, reasonably close to the observed cut-off, suggesting the efficiency of $\eta \approx 0.2$ is justified.

Regarding the pressure anisotropy,  the standard diffusion approximation predicts $\Delta P_{\rm cr}/P_{\rm cr} = 2 (u_{\rm sh}/c)^2$ \citep{JokipiiWilliams}. However, as shown by \cite{LuoMelrose}, the non-resonant short wavelength instability of \cite{Bell04} leads to an \emph{enhancement} of the anisotropy, and requires resonant scattering to counteract it. It is thus desirable to explore the relevant time-scales involved using the updated constraints in this object.

We first consider the case relevant to confined particles that occupy the precursor in the immediate upstream of the shock. {As shown above, the acceleration efficiency, as inferred by the joint Fermi/VERITAS observations, is close to 20\%, unless of course the CRs are mostly confined deep inside the SNR. We thus explore the possible parameter space, keeping fixed the inferred spectral index of $s=4.17$.}
In figures \ref{fig:1} and \ref{fig:2} we plot the phase velocity and growth rates for the full dispersion relation, truncating the expansion at second order. In figure \ref{fig:1} the CR efficiency is kept fixed at the upper limit inferred from observations, and we explore the effect of increasing anisotropy. The first thing to note is that the anisotropy has no effect on the maximum growth rate of the Bell instability
\eqb
\label{maxrate}
\Gamma_{\rm max} = \frac{1}{2}\zeta_1 \frac{v_{\rm A}}{r_{\rm g1}}~.
\eqe
This results from the non-resonant nature of the instability, for which only the total current matters. The CR current is kept fixed for all plots in Fig.\ref{fig:1}. Second, and contrary to the standard picture, only at extremely short wavelengths (or small $\zeta_1$) does the usual Alfv\'en branch emerge. At all other wave-lengths, the modes are highly super-Alfv\'enic and also dispersive. There is thus no unique wave-frame, and naively, one might anticipate this to have an effect on the transport in the precursor, and as a consequence the acceleration. The dispersion free ($\omega/k =$ constant) behaviour that is commonly assumed strictly speaking only applies to $s=4$ power-law spectra, or in the $\zeta_1\ll 1$ limit, where the waves are Alfv\'enic. 
The imaginary components are shown on the right, where we similarly scale by $k$ to highlight the different growth rates for left and right circularly-polarised modes. The blue lines in Figures \ref{fig:1} and \ref{fig:2} show the same data. We do not explore the effect of wave dispersion further here, as the extremely short time-scales associated to the non-resonant waves is likely to dominate the field growth and subsequent particle transport \cite[e.g.][]{Reville13,Bell13}.

\begin{figure*}

	\begin{center}
		\includegraphics[width=0.9\textwidth]{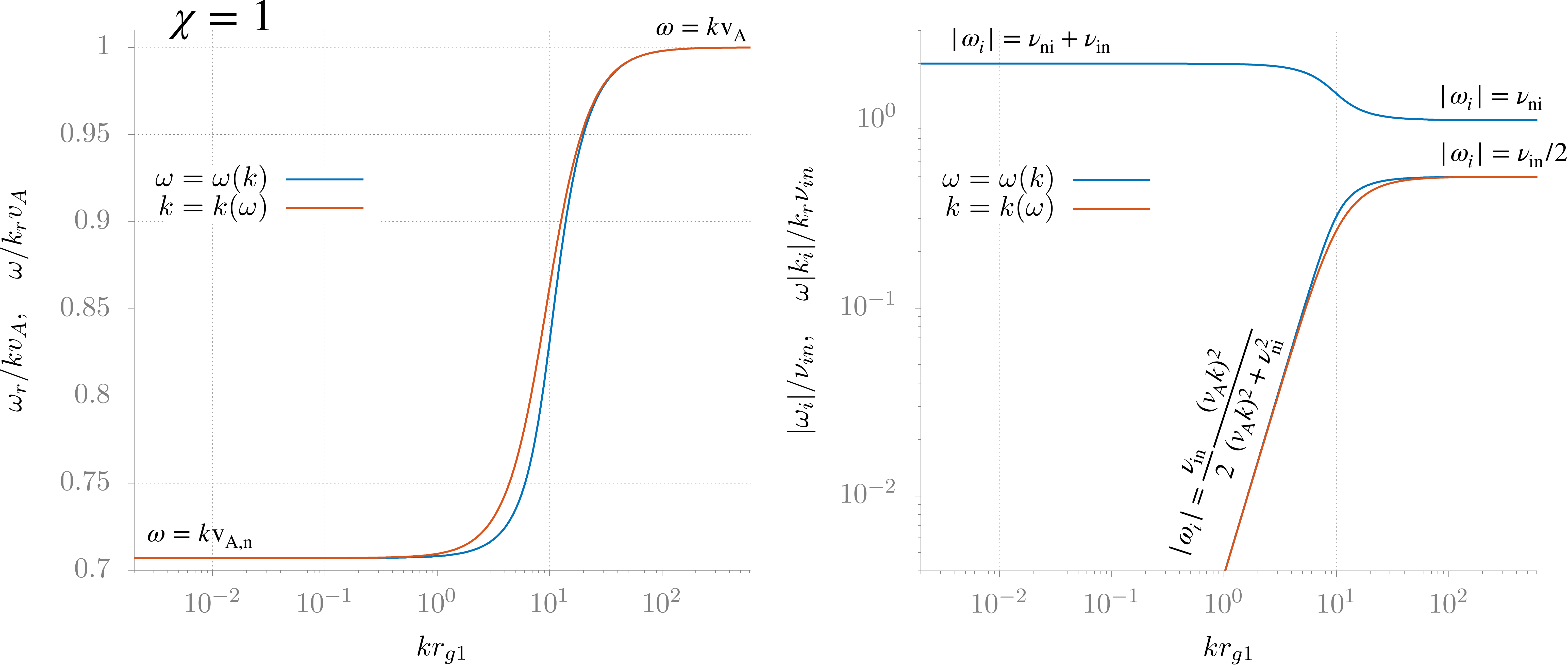} 
			\end{center}
	
~~~~~\line(1,0){450}\\
			\medskip
		\begin{center}
\includegraphics[width=0.9\textwidth]{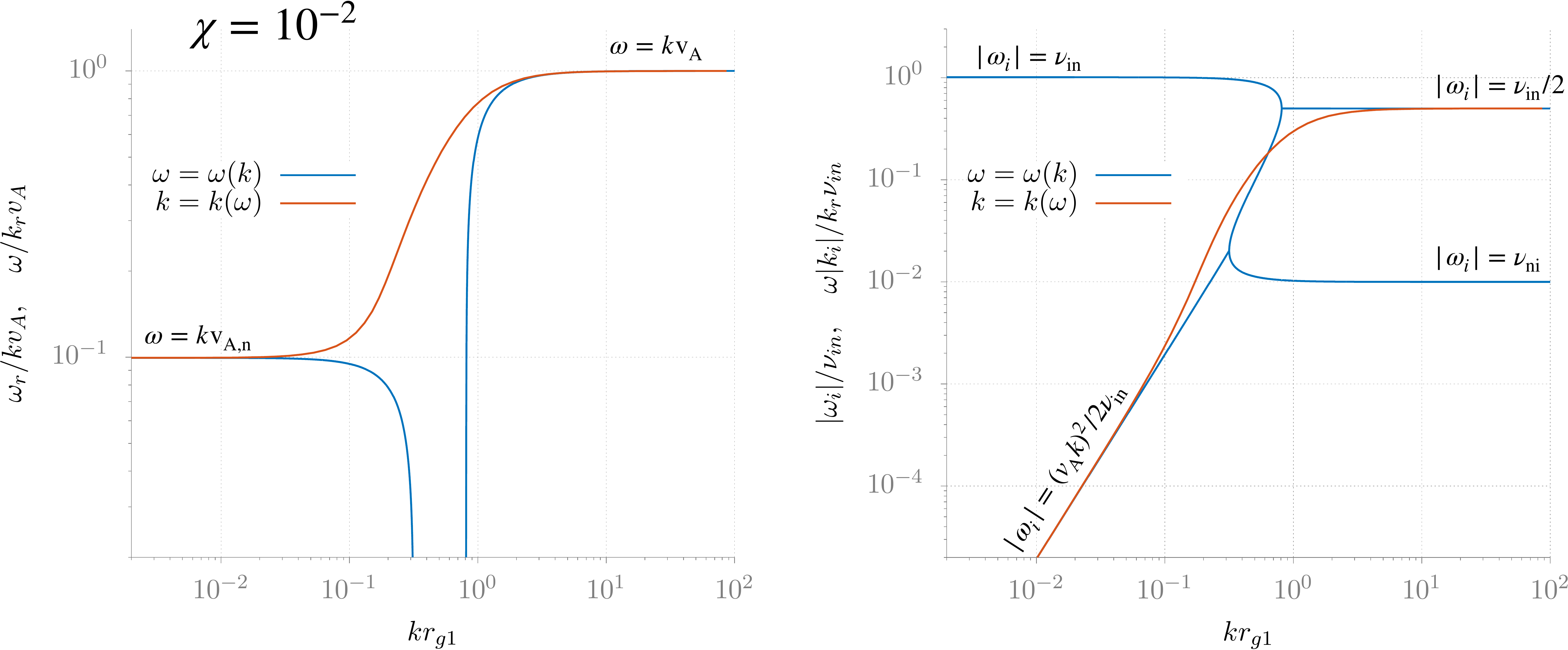}
	\end{center}
	
	\caption{Real ($\omega_r$) and imaginary ($\omega_i$) components of the frequency showing phase velocity and damping rates in the absence of cosmic rays for $\chi=1$ (top) and $\chi=10^{-2}$ (bottom). Both plots assume a fixed $n_{\rm tot} = n_{\rm i}+n_{\rm n} = 10$ cm$^{-3}$, a temperature of $1$~eV, a magnetic field of $5~\mu$G, while $r_{\rm g1}$ corresponds to that of a $100$ GeV proton in this field. In these units, we find for $\chi=10^{-2}$, the evanescent band occurs in the range $0.3 < k r_{\rm g1} < 0.8$, as shown.}
	\label{fig3}
\end{figure*}

In figure \ref{fig:2}, we show the dependence of the phase velocity and growth rates on the cosmic-ray efficiency $\eta$ relative to the reference value $\eta_0=0.2$. This demonstrates the sensitivity of the growth rate, and phase speeds to the assumed cosmic-ray efficiency, as well as the effect of a radial decrease in $\zeta_1$. 
As expected, provided $\zeta_1>1$, scatterers in the resonant band are not adequately described by Alfv\'en waves, as is often assumed to be the case. 
{ Thus, in contrast to the standard model, the range over which the scattering fluctuations are non-Alfv\'enic may extend over a significantly larger volume than 
previously considered. Recently, several groups have performed kinetic simulations of the resonant ion-cyclotron instability in the weak-damping limit \cite[e.g][]{Holcomb,Bai19}. We will explore this process, and its impact on the surrounding environment, in a separate numerical study. }

For Cas A parameters, the reference time $r_{\rm g1}/v_{\rm A}$ is on the order of a few weeks, while the inverse lifetime of Cas A is $r_{\rm g1}/v_{\rm A} t_{\rm SNR} \approx 10^{-4}$. This coincidentally is close to the growth rate for waves resonant with particles close to the inferred cut-off in Cas A, provided acceleration remains efficient. Long wavelength, non-resonant firehose unstable fluctuations however appear to be too slow to be of significance, although an enhancement by almost an order of magnitude could be achieved if the anisotropy is greatly enhanced relative to its diffusive limit (see Fig \ref{fig:1}). The existence of such strong anisotropy would need to be properly motivated.

\section{Cosmic-ray propagation in partially/weakly ionised plasmas}
\label{sec:4}

The linear theory of parallel propagating plane-waves in incompletely ionised magnetised plasmas is well known \citep{KulsrudPearce,ZweibelShull, Tagger95,Xuetal16}, as is its effect on diffusive shock acceleration \citep{Drury96}. Interest around this topic has grown in recent years, primarily in the context of escape from SNRs \citep{Bykov13,Nava19,Inoue}, but also regarding the penetration of CRs into molecular clouds including their role as an ionising agent, as well as their transport in the multi-phase ISM \citep{Padovani,Phan,Brahimi20}. 
{
Complexities associated to the field topology and its impact on CR observable such as the Boron to Carbon ratio have also been explored \citep{DAngelo}.
}

The analysis here follows the approach of \citet{Reville07}. The standard QLT approach of applying the diffusion approximation of \cite{Skilling71}: $\partial f/\partial \mu \approx \lambda_{\rm mfp} \partial f_0/\partial x$ ($x$ here is distance along the mean field) only holds if gradients are long relative to the mean free path $\lambda_{\rm mfp}$. This may not be satisfied for dynamically evolving systems, and hence we keep $\zeta_1$ as a free a parameter. Our aim here is partly to bring attention to areas where 
more numerical efforts are required to address this problem self-consistently.

\subsubsection*{Incompletely ($\chi \sim 1$) and weakly ionised gases without cosmic rays  }

For incompletely ionised ($\chi > 1$) gases, the linear analysis is well known, and we refer the reader to  \cite{ZweibelShull} for a clear presentation. The key limiting results can be read off Fig. \ref{fig3}. At high frequencies, $\omega \gg \nu_{\rm in}$, inter-species collisions do not much affect the oscillations. This results in slowly damped  Alfv\'en waves. At low frequencies $\omega \ll \nu_{\rm ni}$, the ions and neutrals are collisionally coupled, {which results in the two species oscillating as a single fluid with increased inertia. This has the effect of reducing the Alfv\'en  velocity $\omega/k= v_{\rm A,n}$, and since the coupling is more effective at lower frequencies, the damping naturally decreases rapidly ($\propto k^2$) as one moves to longer wavelengths.} The upper blue line in the right hand panel corresponds to damping rates of non-propagating perturbations in the neutral fluid, should they exist.

The case of weakly ionised gases ($\chi \ll 1$) reveals a number of additional complexities, and is of course more relevant to CRs impacting on dense molecular material. In the absence of CRs, and subject ot the condition $\chi < 1/8$, waves in the so-called evanescent range $2 \nu_{\rm ni}/v_{\rm A,n} < k < \nu_{\rm in}/2 v_{\rm A}$ do not propagate \citep{ZweibelShull}.
For sharp resonance ($k=1/r_{\rm g}$) this equates to cosmic-ray energies in the approximate range
\eqb
\label{EvRange}
0.4 \frac{ B_{5\mu {\rm G}}^2}{ n_{\rm i}^{1/2} n_{\rm n} T_{\rm i,eV}^{0.4}}
 < E_{\rm TeV} <
 15 \frac{ B_{5\mu {\rm G}}^2}{\chi_{0.01}^{1/2} n_{\rm i}^{1/2} n_{\rm n} T_{\rm i,eV}^{0.4} }~,
\eqe 
with reference values given in subscripts. For many relevant parameters, this falls in an energy range pertinent for GeV-TeV $\gamma$-ray observations. 
\cite{Tagger95} have pointed out that the wave propagation can be viewed in two ways \cite[see also][]{Soler}. In the generally standard approach of taking coherent plane waves with real $k$, no solutions for the real part of $\omega$ are found in this wave band. However, taking the opposite approach by considering a wave-packet at fixed real $\omega$ and examine how a wave grows/decays as it propagates, \cite{Tagger95} show that waves in this band do in fact propagate, damping as they do so. This latter approach is also well-suited to shock precursor studies, where wave amplitudes are expected to develop in-homogeneously as they approach the shock.
A comparison of the two approaches is shown in Fig. \ref{fig3}, where the lower panels are for an ionisation factor $\chi=10^{-2}$. As expected, the two methods agree at high and low frequencies, but perturbations excited with Fourier components in the evanescent band can clearly be seen to propagate only in the latter approach. Implications are discussed in section \ref{sec:5}

    \begin{figure*}
    	
    	\begin{center}
    	\includegraphics[width=0.95\textwidth]{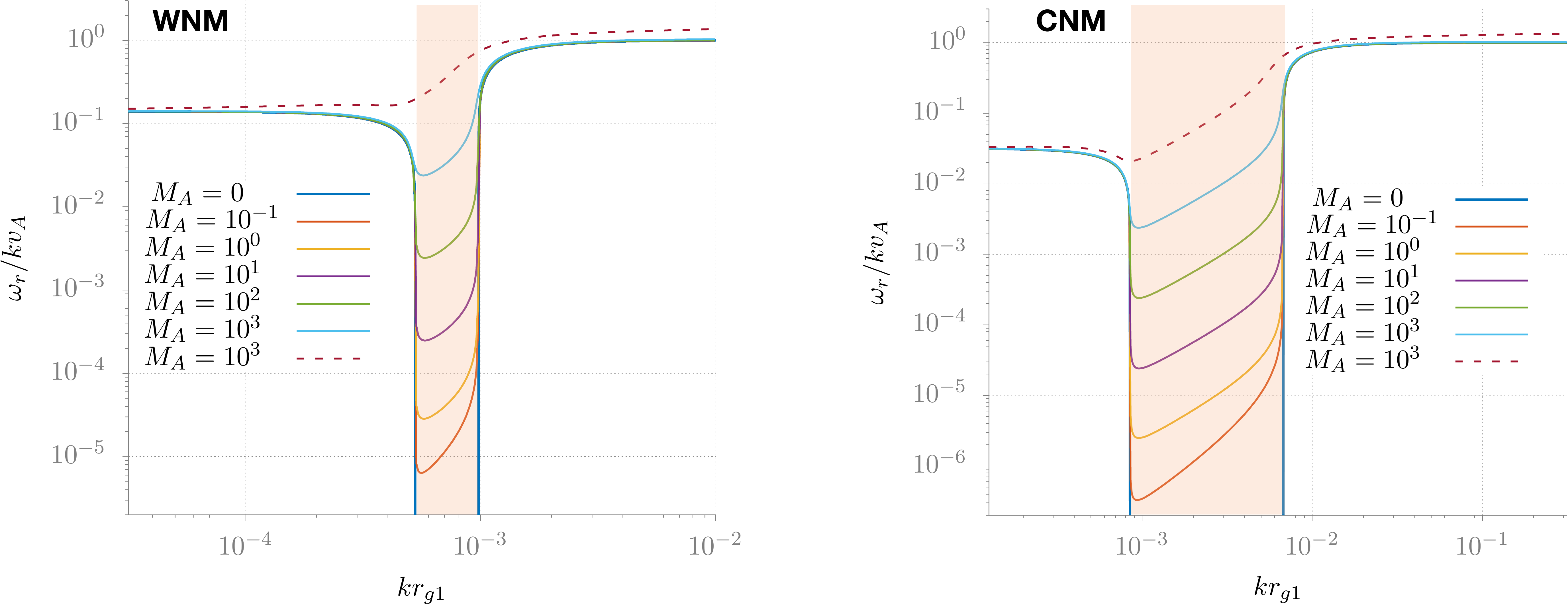} 
    	\end{center}
    	
    	\caption{ Effect of cosmic-rays on evanescent wave-band (shaded-region) for the cases of WNM (left) and CNM (right) as a function of cosmic ray drift speed. Here $M_{\rm A}=v_{\rm cr}/v_{\rm A}$, solid (dashed) lines are for $U_{\rm cr}/U_B =10 (100)$. The cosmic rays are taken to have a spectral index of $s=4.3$ and $E_{\rm min}=1$GeV. 
    	For the WNM we adopt the parameters $n_i+n_n=0.5$ cm$^{-3}$, $T=5,000$K and $\chi^{-1}=50$, while for the 
       CNM we take $n_i+n_n=50$ cm$^{-3}$, $T=50$K and $\chi=10^{-3}$. 
    In both cases the magnetic field is set to $5~\mu$G. Note that for the two dashed lines $\zeta_1>1$. }
    	\label{fig4}
    \end{figure*} 

\subsubsection*{Weakly ionised gases with cosmic rays}

Here, we focus exclusively on gases in which $\chi<1/8$, i.e. the threshold condition for the evanescent band to exist \citep{ZweibelShull}.
The inclusion of CRs into the calculations should be considered with some care. It is convenient to continue working with the cosmic-ray and magnetic energy densities, as both can in principle be inferred from observations. However, one should make sure to check in extremely weakly ionised cases, that the MHD approximation holds, i.e. $n_{\rm cr} \ll n_{\rm i}$. 

The short-wavelength $\zeta_1 > 1$ case was previously considered by \citet{Reville08}.  The present 
results extend those results to arbitrary wavelengths, and upper cut-offs in the distribution. As discussed in the appendix, the contribution from the end points of the distribution must be correctly accounted for to ensure consistent results. 

The damping of scatterers due to ion-neutral collisions in and near CR accelerators has been explored previously in the context of spectral breaks \citep{Malkovetal11}, cosmic ray escape from SNR \cite[e.g.][]{Xuetal16,Brahimi20}, and either the penetration through \citep{CesarskyVolk,Morlino,Inoue} or exclusion from \cite[e.g.][]{SkillingStrong} molecular clouds. Such studies are particularly interesting when they occur in the neighbourhood of a CR source. As discussed above, the existence of an evanescent band is an artefact of the chosen approach, and may not exist in a turbulent plasma, provided the MHD cascade continuously replenishes the wave-band in question. Here we show that even a minute presence of CR anisotropy also causes waves in the evanescent band to propagate.

In Figure \ref{fig4} we consider gas conditions representative of the warm and cold neutral medium phases (WNM and CNM respectively) of the ISM, and the effect of varying CR anisotropy. The details can be read from the caption. Note that we adopted a modest value of $U_{\rm cr}/U_{B} = 10$ in these plots, and explored the effect of increased anisotropy. The dashed curves correspond to $U_{\rm cr}/U_{B} = 100$ to show an example of $\zeta_1>1$ limit. This again shows how waves deviate from standard Alfv\'en branch in the strongly driven regime. While not shown here, it is easily shown that provided $\zeta_1\ll1$, the imaginary part of the frequency matches the expectation from a linear addition of the damping and growth rates (with an appropriate choice for the Alfv\'en velocity at long wavelengths for the latter). This also turns out to be a reasonable approximation to the growth/decay in the  $\zeta_1>1$ limit, i.e. 
\eqb
{\rm Im}(\omega) \approx {\rm Im}(\omega)_{\rm no\, CRs} + G(k) {\rm Im}(\omega)_{\rm no\,neutrals}~,
\eqe
where 
\eqb
G(k)=\left\lbrace\begin{array}{cc}
1 & k> \nu_{\rm in}/2 v_{\rm A} \\ \sqrt{\chi} & k< 2 \sqrt{\chi} \nu_{\rm in}/ v_{\rm A}
\end{array}\right.
\eqe
and with a linear fit in between. This renormalizing factor is necessary to account for the increased inertia of the plasma. The agreement is typically good to within a few percent although slightly larger variations may be found close to the evanescent band that occurs in the absence of CRs.

It is well known that for a fully ionised plasma, the maximum growth rate for field-aligned CR driven modes is
given by the expression in equation (\ref{maxrate}), regardless of the value of $\zeta_1$ \cite[e.g.][]{BellBrazil}. However, there is a fundamental difference between the physical processes underlying the instability in the two limits of $\zeta_1$ greater than or less than unity. The presence of neutrals can expose this difference.   

It is easily verified that the magnitude of the ion-neutral damping rate is always less than the oscillation frequency $ k v_{\rm A}G$. If $\zeta_1>1$, the maximum growth rate is also $ k_{\rm max} v_{\rm A} G$, where $k_{\rm max}=\zeta_1/2r_{\rm g1}$ is the wavenumber of the fastest growing mode. This is merely confirmation of the proof in \cite{Reville07}, that the non-resonant instability can not be stabilised by neutral damping. The opposite is true for resonant modes. If $\zeta_1<1$, and assuming the distribution is peaked at an energy corresponding to gyroradius $r_{\rm g1}$, by definition $k_{\rm max}=1/r_{\rm g1}$ and $\gamma_{\rm max} <  G k v_{\rm A}$ and hence can be stabilised. 

We thus seek roots to the equation  
\eqb
{\rm Im}(\omega)_{\rm no\, CRs} + G(k) {\rm Im}(\omega)_{\rm no\,neutrals}=0~.
\eqe
Two possibilities exist. If $r_{\rm g1} < \zeta_1 v_{\rm A}/\nu_{\rm in}$ (c.f. the lower value in the inequality (\ref{EvRange})) growth is still possible, however for small $\zeta_1$ this may correspond to very low energies. On the other hand, if $r_{\rm g1} > \zeta_1 v_{\rm A}/\nu_{\rm in}$ net damping is expected at short wavelength, but since the damping rate at long wavelengths scales as $k^2$, and resonantly driven modes scale as $k^{s-3}$ (neglecting pressure anisotropy effects), higher energy particles, if present, can still produce net growth.  

Thus it is possible to define a critical energy, below which 
damping must dominate. 
\eqb
r_{\rm g}^{5-s} = \frac{r_{\rm g1}^{4-s}}{\sqrt{\chi}\zeta_1}\frac{v_{\rm A}}{\nu_{in}}~.
\eqe
Taking for simplicity $s=4$, we see this equates to an energy $\zeta_1^{-1}$ times the upper energy limit in the inequality (\ref{EvRange}), with all energies below this being unable to self-generate scattering waves. 
While it is tempting to speculate on the impact of this effect for escape from SNR, or indeed penetration into a molecular cloud, we will defer such discussion until reliable numerical simulations can explore the full multi-scale aspects.  

\section{Discussion}
\label{sec:5}

The stability of parallel propagating transverse MHD fluctuations in the presence of an external current of cosmic rays has been explored. The solution given in equation (\ref{DispMaster}), and in particular the method for determining the cosmic-ray response provided in the appendix, generalises many of the existing results in the literature. The approach can readily be applied to other plasma conditions, where parallel propagating modes are of interest. A limitation of the analytic results presented here, is the use of a single power-law dependence at all orders of anisotropy. For example, in the context of shock precursors, one might anticipate the anisotropy at large momenta to have a different spatial profile to that at low momenta, which might for example enhance the growth rate of the fire-hose instability. A follow-up study of cosmic-ray anisotropy in multi-scale self-generated fields will exploit the generic form presented here.

The standard quasi-linear theory approach to studying cosmic-ray transport and self-confinement applies in the limit of $\zeta_1<1$, or $U_{\rm cr}/U_B \ll c/v_{\rm cr}$. In the Galactic disk, we expect $U_{\rm cr} \approx U_B$ and $v_{\rm cr} \approx v_{\rm A} \ll c$, meaning this condition is generally satisfied. However, near to sources, where both $U_{\rm cr}$ and $v_{\rm cr}$ are greatly enhanced, this is generally not the case. 
Similar conclusions may apply for the low density environment at the periphery of the Galactic halo \cite[e.g.][]{BlasiAmato}. 
We plan to bridge the two regimes in an upcoming numerical study. 

The transport of cosmic-rays in the ISM, and the related topic of their  escape from sources, has recently received significant interest, with particular emphasis on the different phases of the ISM. The role of ion-neutral damping features heavily in these studies. That different conceptual approaches to the problem 
produce opposite conclusions regarding propagation of Alfv\'enic fluctuations at intermediate frequencies in weakly ionised plasmas ($\nu_{\rm ni} < 
\omega< \nu_{\rm in}$) is already known \citep{Tagger95,Soler}.  However, studies to date have focussed exclusively on only one such approach, which, motivated by the new results, we now argue is in fact less likely to apply. 

Turbulent MHD cascades can be conceptualised as interactions between colliding wave-packets \citep{Kraichnan}. To mediate efficient exchange of energy between Fourier modes, the semi-classical conservation laws for three-wave couplings: $\bm{k_1}+\bm{k_2}= \bm{k_3}$ and $\omega(\bm{k_1})+\omega(\bm{k_2})= \omega(\bm{k_3})$ should be satisfied.  
Here $\omega(\bm{k})$ is the 3D generalisation of the dispersion relation in this paper \footnote{ \citet{Bell2005} provides the result without neutrals for fixed CR current, while \citet{Soler} provides the result for neutrals without CRs.}. It is well-known that for incompressible MHD, coupling requires either vanishing $\bm{k}_1\cdot \bm{v}_{\rm A}$ or $\bm{k}_2\cdot \bm{v}_{\rm A}$, i.e. one of the Fourier modes must be non-propagating and in a plane perpendicular to the guide field. This generally results in an anisotropic turbulence spectrum \cite[see][for a review]{Zhou}. If the waves are however dispersive, which will always be the case in the neighbourhood of an hypothetical evanescent band, it is straightforward to show that this is no longer a requirement, provided $\omega \neq 0$. 
Thus, based on the dispersion curves shown in section \ref{sec:4}, there is no physical justification to exclude waves with $\bm{k_3}$ falling within the evanescent band, and allowing them to propagate with phase velocity between the two Alfv\'en velocity limits, as shown in Fig \ref{fig3}. The existence of a region of non-propagating fluctuations would appear to be a highly idealised concept. The dispersive nature of the waves however may still cause modifications to the resulting turbulence spectrum and may yet leave an imprint on energetic particle transport. To our knowledge, there has been no thorough investigation of this effect, which may reveal itself in the $\gamma$-ray emission from CRs in molecular clouds.

To this end, recent studies of the diffuse $\gamma$-ray emission from giant molecular clouds with Fermi-LAT \cite[e.g.][]{Aharonian} hint at a dependence of the Galactic cosmic-ray density with Galacto-centric radius, although could also correlate with nearby sources. The lack of any clear trend in the spectra however hint at the absence of unique energy-dependent behaviour/trends. Surveys that look at higher energies with instruments such as HAWC \citep{Lauer}, LHAASO \citep{LHAASO} and the anticipated southern hemisphere counterpart SWGO \citep{SWGO}, may provide a different picture.  The analysis presented in this work is important for the transport at such energies, and future observations in the VHE to UHE  $\gamma$-ray regimes may reveal new insight into the role of cosmic-ray feedback on molecular clouds. Unfortunately, we currently lack predictive capability.

Finally, we note that attention was restricted in the present study to relativistic particles. The results could however be extended to non-relativistic particles without difficulty. Since low energy cosmic rays may play an important role in the ionisation of dense molecular clouds, the ability of self-induced scattering to inhibit penetration into the cores of such clouds warrants further investigation.

\section*{Acknowledgements}

The authors wish to thank J. Kirk and H. Voelk for comments and discussion. BR acknowledges conversations with attendees of the Multiscale Phenomena in Plasma Astrophysics workshop at the Kavli Institute for Theoretical Physics. This research was supported in part by the National Science Foundation under Grant No. NSF PHY-1748958.

\section*{DATA AVAILABILITY}

No new data were generated or analysed in support of this research.



\bibliographystyle{mnras}


\onecolumn 
\appendix

{
\section{Dispersion relation -kinetic  vs fluid approach}
}
For completeness, we compare different approaches to deriving the linear dispersion relation
for circularly-polarized modes propagating parallel to a background field ($\bm{B}_0= B_0 \bm{\hat{x}}$). Following standard text-books, we explore first order small perturbations as a spectrum of periodic transverse waves \cite[e.g.][]{Ichimaru,KrallTrivelpiece}. For a uniform background, the zeroth order total current and charge density must vanish.  
Taking all first order quantities to be of the form $\xi(x,t) = \xi_k {\rm e}^{i(k x-\omega t)}$, with $\bm{k\cdot \delta E}_k = 0$ Maxwell's equations can be combined to give
\eqb
\label{EqA1}
(\omega^2 - c^2k^2) \bm{\delta E}_k + 4 \pi i \omega \sum_s \bm{j}_{\bot,s} = 0 ~,
\eqe
where 
\eqb
\bm{j}_{\bot,s} = q_s\int \delta f_{k,s} \bm {v}_\bot d^3 p~,
\eqe
is the perturbed current associated to species $s$, with
$q_s$ its charge. The summation is taken over all species in the plasma.
 
To derive a dispersion relation, it is necessary to describe the response of the current to a fluctuating magnetic field. This is determined from the Vlasov equation, using the zero-th order helical trajectories along the mean field\cite[see for example][chapter 8]{KrallTrivelpiece}, and gives
\eqb
\bm{j}_{\bot,s} =
i\frac{q_s^2}{2} \sum_{n=\pm 1} \int d^3 p
\frac{v_\bot p_\bot}{\omega  - n \omega_{{\rm g},s} -k v_\| +{\rm i}0^+}
\left[ 
\frac{ k v_\|}{\omega}\left(\frac{\partial f_s}{\partial p_\bot^2}-\frac{\partial f_s}{\partial p_\|^2}\right) -\frac{\partial f_s}{\partial p_\bot^2} \right] 
\left(\begin{array}{cc}
1 & i n \\ -i n & 1
\end{array}\right) \bm{\delta E}_k~.
\eqe
Here $f_s(p_\|,p_\bot^2)$ is the gyrotropic zeroth order equilibrium solution of sepcies $s$. The ${\rm i}0^+$ term ensures the integration is taken along the Landau contour, and $\omega_{{\rm g},s}=q_sB_0/\gamma m_s c$ is the relativistic gyro frequency.
Introducing the complex quantities $j_\bot = j_y +{\rm i} \epsilon j_z$ (and similarly for $\delta E$), where $\epsilon = \pm1$ determines the polarisation, the current takes the rather simple form
\eqb
j_{\bot,s} =
{\rm i} {q_s^2} \int d^3 p
\frac{v_\bot p_\bot}{\omega  -  \epsilon \omega_{{\rm g},s} -k v_\| +{\rm i}0^+}
\left[ 
\frac{ k v_\|}{\omega}\left(\frac{\partial f_s}{\partial p_\bot^2}-\frac{\partial f_s}{\partial p_\|^2}\right) -\frac{\partial f_s}{\partial p_\bot^2} \right] 
{\delta E}_k~.
\label{jbot}
\eqe 
This expression is found in most textbooks, however, it proves useful to write it out explicitly here, as we will use it in the next section.

Applying the same notation to Eq (\ref{EqA1}) and substituting in the expression for the current, one recovers the familiar expression
\eqb
\frac{c^2k^2}{\omega^2} - 1 = \sum_s \mathcal{X}_s
\eqe
where the plasma susceptibility 
\eqb
\mathcal{X}_s =  \frac{4 \pi q_s^2}{\omega} \int d^3 p
\frac{v_\bot p_\bot}{\omega  -  \epsilon \omega_{{\rm g},s} -k v_\| +{\rm i}0^+}
\left[ \frac{\partial f_s}{\partial p_\bot^2}+
\frac{ k v_\|}{\omega}\left(\frac{\partial f_s}{\partial p_\|^2}-\frac{\partial f_s}{\partial p_\bot^2}\right)
 \right] 
 \label{EqSusc1}
\eqe
has been introduced. 

This is the starting point for kinetic investigations of circularly polarised linear modes that propagate parallel to a guide field. As is standard, an equilibrium condition must be considered that satisfies the general constraints of zero net charge and current. When a current is provided by streaming cosmic rays, a compensating return current must be drawn by the background plasma. Since the cosmic rays are generally small by way of number, one approach is to simply assume the thermal electrons establish a slow drift relative to the background ions that provides the required return current \cite[e.g.][]{Achterberg83, Reville06,LuoMelrose}.  An alternative scenario involves only a sub-population of electrons are drawn to provide the return current \citep{AmatoBlasi,ZweibelEverett}.
We note however, that such a configuration is also two-stream unstable, and for realistic scenarios, is likely to relax rapidly to the former. In the work of \citet{AmatoBlasi} the two  approaches are shown to ultimately give equivalent results, at least in the low frequency regime of interest. In this regime, it is of course expected that the results approach the MHD limit, which is insensitive to  the microphysics of the thermal gas.

In the fluid approach, one need only consider the bulk properties of all quantities, although a general treatment still requires a kinetic treatment of the CRs. The fluid approach has the tremendous advantage that it allows easy inclusion of collisions, something that would be undesirably cumbersome in a kinetic approach.

Starting from the fluid equations for the thermal species, which for simplicity  we limit to be electrons, ions and neutrals, for each species we have
\eqb
n_s m_s \frac{d\bm{u}_s}{dt} = -\nabla P_s + n_s q_s\left(\bm{E} + \frac{1}{c}\bm{u}_s\times \bm{B}\right) +  n_s m_s  \sum_{\alpha\neq s} \nu_{s\alpha}(\bm{u}_s-\bm{u}_\alpha)
\eqe
where $\nu_{s\alpha}$ is the interspecies momentum exchange rate, and we have assumed isotropic pressure for each species. In the following, we consider only collisions between the ions and neutrals. 

We proceed by summing over the charged components in the usual fashion, to give
\eqb
\rho \frac{d\bm{u}}{dt} = -\nabla P_{\rm th} +  \rho_i \nu_{in}(\bm{u}_i-\bm{u}_n)+\left(\sum_{s=i,e}n_s q_s\right)\bm{E} + \frac{1}{c}\left(\sum_{s=i,e}n_s q_s \bm{u}_s\right)\times \bm{B} 
\eqe
where $\rho = n_i m_i + n_e m_e \approx \rho_i$ and $\bm{u} = (n_i m_i \bm{u}_i+ n_e m_e \bm{u}_e)/\rho \approx  \bm{u}_i$ . In the absence of CRs, the last two terms equate to $0$ and $\bm{j}_{tot}$ respectively. However, as in the kinetic description above, we must not neglect the CR contribution. Thus, from quasineutrality and Ampere's law, we have
\eqb
\sum_{s=i,e}n_s q_s = -n_{\rm cr} q_{\rm cr} \mbox{~~~and~~~}
 \frac{c}{4 \pi} \bm{\nabla}\times \bm{B} = \bm{j}_{\rm cr}+\sum_{s=i,e}n_s q_s \bm{u}_s 
 \eqe
Subsituting into the above, we find
\eqb
\rho_i \frac{d\bm{u_i}}{dt} = -\nabla P_{\rm th} +  \rho_i \nu_{in}(\bm{u}_i-\bm{u}_n) - n_{\rm cr} q_{\rm cr} \bm{E} + \frac{1}{c}\left(  \frac{c}{4 \pi} \bm{\nabla}\times \bm{B} - \bm{j}_{\rm cr} \right)\times \bm{B} 
\eqe
which in the ideal MHD limit, $\bm{E} = -\bm{u}\times\bm{B}/c$ reduces to Equation (\ref{momEqn}). 

The kinetic elements of this problem are introduced via $\bm{j}_{\rm cr}$, which we describe in the next section.

\section{The cosmic-ray magnetic response}

Following \cite{Bell04}, we wish to express the response of the cosmic-ray current to fluctuations in the magnetic field. As discussed and summarised above, this equivalent procedure for electric fields can be found in most plasma text books. However, we can easily connect the two methods.
Using the complex notation from the previous section, it follows from Faradays' law, 
\eqb
\delta E = -{\rm i}\epsilon \frac{\omega}{ck}  B_\bot~,
\eqe
which on substitution into (\ref{jbot}) gives
\begin{equation*}
j_\bot = \epsilon q^2\int d^3 p \frac{p_\bot v_\bot}{\omega - \epsilon\omega_g - k v_\| + {\rm i}0^+}
\left[ \frac{v_\|}{c}\left(\frac{\partial f}{\partial p_\bot^2}-\frac{\partial f}{\partial p_\|^2}\right)
-\frac{\omega}{ck}\frac{\partial f}{\partial p_\bot^2}\right] B_\bot~.
\end{equation*}
Note we have dropped the subscript for species, as we are only concerned with CRs.

We seek a general formalism to determine the response for a given Legendre polynomial expansion of the distribution function. In this case it is convenient to first transform to spherical momentum coordinates 
\begin{equation}
\label{jperp}
j_\bot = \epsilon \frac{\pi q^2}{k}\int dp~ p \int d\mu  \frac{1-\mu^2}{\mu - \mu_0 -  {\rm i}0^+}
\left[ \frac{v}{c}\frac{\partial f}{\partial \mu}
+\frac{\omega}{ck} \left( p \frac{\partial f}{\partial p}- \mu\frac{\partial f}{\partial \mu} \right) \right] B_\bot~,
\end{equation}
where we have introduced $\mu_0 = \frac{\omega}{kv}-\epsilon\frac{\omega_g}{kv}$, with $k>0$.

Expansion of the distribution in a Legendre polynomial series $f(p,\mu) = \sum f_\ell(p) P_\ell(\mu)$, leads to
\eqnb
j_\bot =\frac{  4 \pi q^2}{3k}  \left[ \sum_{n=0}^\infty \int dp~ p  A_n I_n \right] B_\bot~,
\eqne
where from Rodrigues' formula one can show
\begin{align*}
&A_n =  \frac{v}{c} (n+1) \sum_{\ell=n+1}^{\infty}  a^\ell_{n+1} {f_\ell}+
\frac{\omega}{ck}\sum_{\ell=n}^{\infty}  a^\ell_{n}\left(
 p\frac{\partial}{\partial p} - n \right)f_\ell~,
\end{align*}
with generalised binomial coefficients 
\eqnb
 a^\ell_{n} = 2^\ell \binom{\ell}{n} \binom{\frac{\ell +n-1}{2}}{\ell} ~.
\eqne 
The integration over pitch angle is contained in the term
\eqnb
 I_n =\frac{3 \epsilon}{4} \int_{-1}^1 d\mu \frac{(1-\mu^2)\mu^n} {\mu-\mu_0 -  {\rm i}0^+}\enspace.
 \eqne 
 
\noindent
Recall that $j_\bot/j_{0} = \sigma B_\bot/B_0$ and therefore
\eqnb
\sigma =\frac{  4 \pi q^2}{3k}  \frac{B_0}{j_0} \sum_{n=0}^\infty \int dp~ p  A_n I_n  ~.
\eqne
This format is convenient when the series is to be truncated after a finite number of terms, i.e. setting $f_{\ell > L_{\rm max}}=0$.
Consider for example $L_{\rm max}=3$, the first few $A_n$ terms are
\begin{align*}
&A_0 = \frac{v}{c} \left(f_1-\frac{3}{2}f_3\right) +\frac{\omega}{ck}p\frac{\partial ~}{\partial p}\left(f_0 - \frac{1}{2} f_2\right)~,\nonumber\\
&A_1 =3 \frac{v}{c} f_2 +\frac{\omega}{ck}\left(p\frac{\partial}{\partial p}- 1\right)
\left(f_1-\frac{3}{2} f_3\right) ~, \\
&A_2 = \frac{15}{2} \frac{v}{c} f_3 +\frac{3}{2}\frac{\omega}{ck}\left(p\frac{\partial f_2}{\partial p}- 2 f_2\right)~,\nonumber \\
&A_3 = \frac{5}{2}\frac{\omega}{ck}\left(p\frac{\partial f_3}{\partial p}- 3 f_3\right)~.
\end{align*}

For the integral over $\mu$ in $I_n$, we note that if $|\mu_0|<1$, the integral has a simple pole at $\mu=\mu_0$, and hence in general we may write
\eqnb 
 I_n=  \frac{3 \epsilon}{4} \left \lbrace  {\rm i }\pi \left[ 1-{\rm Min}(1,\mu_0^2)\right]\mu_0^n +\mathcal{ P} \int_{-1}^1 d\mu \frac{(1-\mu^2)\mu^n} {\mu-\mu_0 } 
  \right\rbrace ~,
 \label{Eq:In}
\eqne
with $\mathcal{P}$ denoting the Cauchy principal value. The $I_n$ satisfy the recurrence relation
\eqnb
I_{n+1} = \mu_0 I_n + \frac{3  \epsilon}{2} \frac{1+\cos n\pi}{(n+1)(n+3)}
\eqne
and hence in computing the pitch angle integration it is necessary to specify only $I_0$:
\eqnb
I_0 = \frac{3 \epsilon}{4} \left \lbrace
\left(1-\mu_0^2\right)\ln\left|\frac{1-\mu_0}{1+\mu_0} \right|
-2\mu_0  + {\rm i }\pi \left[ 1-{\rm Min}(1,\mu_0^2)\right]\right\rbrace
~.
\eqne

This formalism is still completely general (we have not made any assumptions about $\omega$ in high/low frequency limits, etc.). It is particularly convenient for numerical calculations, as the final integral over momentum can be performed using standard methods. The approach may thus be applied to other scenarios where field aligned modes are of interest.
A similar result is found using Chebyshev polynomials for the angular basis functions, but since one of the motivations is to provide a general framework for comparison with future simulations using an expansion in the related Spherical Harmonic basis functions \citep{Reville13}, for brevity, we leave this extension to the reader.

Before proceeding to derive expressions for power-law distributions of relativistic particles, we make a brief digression to demonstrate how the above technique could be applied to other cases, such as a slow drifting Maxwellian distributions. To compare with previous results, we use the previous expression for the susceptibility,  Equation (\ref{EqSusc1}), which in spherical coordinates reads
\eqnb
\mathcal{X}_s = -\frac{4 \pi^2 q^2 c}{ \omega^2} \int dp ~ p \int d\mu  \frac{1-\mu^2}{\mu - \mu_0 -  {\rm i}0^+}
\left[ \frac{v}{c}\frac{\partial f}{\partial \mu}
+\frac{\omega}{ck} \left( p \frac{\partial f}{\partial p}- \mu\frac{\partial f}{\partial \mu} \right) \right]~.
\eqne
Noting the similarity to equation (\ref{jperp}), the same expansion approach applies, and we find
\eqnb
\mathcal{X}_s =-\varepsilon\frac{  16 \pi^2 q^2 c}{3 \omega^2}  \left[ \sum_{n=0}^\infty \int dp~ p  A_n I_n \right] ~.
\eqne

Consider a Maxwell-Boltzmann distribution of particles with mass $m$,  temperature $T$ and a non-relativistic drift velocity $\bm{v}_d = v_d \bm{\hat{x}}$:
\eqnb
f_{MB}(\bm{p}) = \frac{n}{(2 \pi p_{th}^2)^{3/2}} {\rm e}^{-(\bm{p}-\bm{p}_d)^2/2 p_{th}^2}
\eqne
where $p_{th} = \sqrt{m k_B T}$, and $\bm{p}_d=m \bm{v}_d$. As above, the distribution is expanded in Legendre polynomials, 
\eqnb
f_\ell = \frac{2 \ell +1}{2}  \frac{n}{(2 \pi p_{th}^2)^{3/2}} {\rm e}^{-\frac{{p}^2+p_d^2}{2 p_{th}^2}} \int_{-1}^{1}  d\mu P_\ell(\mu) ~ \exp\left[{\frac{p p_d}{p_{th}^2} \mu}\right]
\eqne
which provided $p_d \ll p_{th}$, converges rapidly. The first 3 terms are
\begin{align*}
&f_0 =   \frac{p_{th}^2}{2pp_d}\left[\frac{n}{(2 \pi p_{th}^2)^{3/2}}\left( {\rm e}^{-\frac{1}{2} \left( \frac{p-p_d}{p_{th}}\right)^2 }
-{\rm e}^{-\frac{1}{2} \left( \frac{p+p_d}{p_{th}}\right)^2}\right)\right] ~\approx~ 
\frac{n}{(2 \pi p_{th}^2)^{3/2}} {\rm e}^{-p^2/2 p_{th}^2}
\left[1+\frac{p_d^2}{ p_{th}^2}\frac{p^2-3 p_{th}^2}{6 p_{th}^2}\right]
\\
&f_1=   \frac{3 p_{th}^2}{2pp_d}\left[\frac{n}{(2 \pi p_{th}^2)^{3/2}}\left( {\rm e}^{-\frac{1}{2} \left( \frac{p-p_d}{p_{th}}\right)^2 }
+{\rm e}^{-\frac{1}{2} \left( \frac{p+p_d}{p_{th}}\right)^2}\right) ~ - ~2 f_0\right] ~\approx~ 
\frac{n}{(2 \pi p_{th}^2)^{3/2}} {\rm e}^{-p^2/2 p_{th}^2}
\left[\frac{p_d p}{ p_{th}^2}\right]\\
&f_2=5\left[ f_0 -  \frac{p_{th}^2}{pp_d} f_1\right]  ~\approx~ 
\frac{n}{(2 \pi p_{th}^2)^{3/2}} {\rm e}^{-p^2/2 p_{th}^2}
\left[\frac{p^2_d p^2}{3 p_{th}^4}\right]
\end{align*}
where the approximate terms are accurate to $\mathcal{O} \left[(p_d/p_{th})^2\right]$.

Keeping only terms to first order in $p_d/p_{th}$, it follows that for wavelengths $k v_{th}/\omega_{\rm g} \ll 1$, 
\eqnb
\mathcal{X}_s = \frac{\omega_{\rm p}^2 (\omega -k v_d)}{\omega^2 (\omega-\varepsilon \omega_{\rm g})}
\left[ {\rm i} \sqrt{\frac{\pi}{2}} \mu_0 {\rm e}^{-Z^2/2} - \frac{1}{Z^2} -1\right]
\eqne
where  $\omega_{\rm p} = (4 \pi n q^2/m)^{1/2}$ is the plasma frequency, and $Z=\mu_0|_{v=v_{th}}$. This agrees with the standard result for the long wavelength limit of the plasma dispersion function \cite[e.g.][]{Ichimaru}. While the approach is not as straightforward, the algorithm is simple and trivial to implement numerically. Most importantly, it may prove useful for situations in which the moments of the distribution function do not reduce to standard integrals.

\section{Solving for power-law Cosmic Ray distributions}

To simplify the analysis that follows (although not strictly necessary at this stage), we consider only modes with $\omega \ll \omega_g$ (which can be checked after the fact) making $\mu_0= - \epsilon \omega_g/kv \equiv - \epsilon\lambda$. Here we have defined the normalised wavelength $\lambda = \omega_g/kv = (k r_g)^{-1}$.
In this limit
\begin{align*}
&I_0 =  \frac{3}{2} \lambda +\frac{3}{4} \left(1-\lambda^2\right)\ln\left|\frac{1+\lambda}{1-\lambda} \right| +  \epsilon \frac{3\pi {\rm i}}{4} [1-{\rm Min}(1,\lambda^2)]\\
 & I_{n+1} =  \epsilon  \left[  \frac{3(1+\cos n\pi)}{2(n+1)(n+3)}-\lambda I_n\right]
\end{align*} 
 
We seek a closed form solution to develop astrophysical applications. To this end, we  follow \cite{Bell04}, taking a power-law distribution $f_\ell(p) = \phi_\ell p^{-s} \Theta(p;p_1,p_2)$ where  $\Theta(p;p_1,p_2)=H(p-p_1) - H(p-p_2)$, 
and $H(p)=\int_{-\infty}^p \delta(x) dx$ is the Heaviside step function.  The assumption that all orders in the expansion have the same power-law shape is an obvious short-coming, as in non-equilibrium systems the anisotropy may well have a non-trivial momentum dependence, but we defer any detailed discussion to future numerical experiments, and a full non-linear theory is beyond the scope of this paper. 

In the following, we consider only situations where $p_1 \gg mc$ such that we can approximate $v/c =1$. We note that 
\begin{equation*}
p\frac{\partial f_\ell}{\partial p}-nf_\ell = \phi_\ell p^{1-s} \left[\delta(p-p_1) - \delta(p-p_2)\right] -(n+s) f_\ell 
\end{equation*}
In their final results, \cite{Bell04} and \cite{Reville07} ignored the contribution from the end point delta-functions, as they play no role in the short wavelength limit. However, they are essential in the long wavelength limit,  and neglecting them would result in the unphysical result of $\lim_{k\rightarrow 0}\sigma
\neq 1$, i.e. cosmic rays would not follow the field lines on large scales!

Substituting into $A_n$, and changing variable to $\lambda = \lambda_1 p_1/p =  \lambda_2 p_2/p$, we find
\begin{align*}
A_n & = \frac{ \lambda^{s-1} }{(p_1 \lambda_1)^s}  \left\lbrace
 (n+1) \sum_{\ell=n+1}^{{L_{\rm max}}}  a^\ell_{n+1} {\phi_\ell}\lambda+
\frac{\omega}{ck}\sum_{\ell=n}^{{L_{\rm max}}}  a^\ell_{n}\phi_\ell
\left[
\lambda_1^2\delta(\lambda-\lambda_1) - \lambda_2^2\delta(\lambda-\lambda_2)
-(n+s)   \lambda  
\right]
\right\rbrace \Theta(\lambda;\lambda_2,\lambda_1) \\
&= \frac{ \lambda^{s-1} }{(p_1 \lambda_1)^s}   \Theta(\lambda;\lambda_2,\lambda_1) \tilde{A}_n
\end{align*}
Likewise, changing variable in the momentum integral it follows that
\eqnb
\sigma =\frac{  4 \pi}{3n_{\rm cr}}  \frac{c}{v_{\rm cr}} (p_1 \lambda_1)^{3-s} 
\sum_{n=0}^{L_{\rm max}} \int_{\lambda_2} ^{\lambda_1} d\lambda~ \lambda^{s-4}  \tilde{A}_n I_n 
\eqne
where we have introduced the bulk CR velocity $v_{\rm cr}= j_{\rm cr}/ q n_{\rm cr}$.
From here on, we restrict our attention to positive $k$ (and $\lambda$) and define the integral
\eqnb
\label{Kint}
K_n^m(\lambda) = \int^\lambda d\zeta ~ \zeta^m I_n(\zeta) ~,
\eqne
which also satisfies a useful recursion relation:
\begin{equation}
\label{Kident}
K_n^{m} + \epsilon K_{n+1}^{m-1}= \frac{3}{2}\frac{1+\cos n\pi}{(n+1)(n+3)}\frac{\lambda^{m}}{m} \enspace .
\end{equation}
Using this allows us to write
\eqnb
\sigma =\frac{  4 \pi}{3n_{\rm cr}}  (p_1 \lambda_1)^{3-s} 
\sum_{n=0}^{L_{\rm max}} 
\left[
 (n+1) \frac{c}{v_{\rm cr}} \sum_{\ell=n+1}^{{L_{\rm max}}}  a^\ell_{n+1} {\phi_\ell}K_n^{s-3}+
\frac{\omega}{v_{\rm cr} k}\sum_{\ell=n}^{{L_{\rm max}}}  a^\ell_{n}\phi_\ell
\left(
\lambda^{s-2} I_n
-(n+s)  K_n^{s-3}
\right)
\right]_{\lambda_2} ^{\lambda_1} 
\eqne
Finally, for convenience in the main part of the text, we will split up the linear susceptibility into frequency dependent/independent parts, 
\begin{equation}
\label{Eq:SigmaFinal}
\sigma =\left[ \sigma_1 -  \frac{\omega}{kv_{\rm cr}} \sigma_2\right]_{\lambda_2} ^{\lambda_1} ,
\end{equation}
where   
\begin{align*}
&\sigma_1 =\frac{  4 \pi}{3n_{\rm cr}}  (p_1 \lambda_1)^{3-s} \frac{c}{v_{\rm cr}} 
\sum_{n=0}^{L_{\rm max}} 
 \sum_{\ell=n+1}^{{L_{\rm max}}}   (n+1)  a^\ell_{n+1} {\phi_\ell}K_n^{s-3}\\
&\sigma_2 =\frac{  4 \pi}{3n_{\rm cr}}  (p_1 \lambda_1)^{3-s} 
\sum_{n=0}^{L_{\rm max}} 
\sum_{\ell=n}^{{L_{\rm max}}}  a^\ell_{n}\phi_\ell
\left[
(n+s)  K_n^{s-3}-\lambda^{s-2} I_n 
\right]\enspace .
\end{align*}

\subsection{Evaluating $K_n^m$}

From the definition of $I_n$ it follows that $K_n^m$ contains only terms which are either integrals of powers of $\lambda$ or the logarithmic function via:
\eqnb
L_m = \int d\lambda ~\lambda^m \left(1-\lambda^2\right)\ln\left|\frac{1+\lambda}{1-\lambda} \right|  \enspace ,
\eqne
where $m\geq 0$. The latter has solution for general $m$
\eqnb
L_m= \left[\frac{\lambda^{1+m}}{1+m}- \frac{\lambda^{3+m}}{3+m}\right]\ln\left|\frac{1+\lambda}{1-\lambda} \right|
+F_{m+2} - F_{m}
\eqne
where Gauss'  Hypergeometric function appears via
\eqnb
F_m(\lambda) = {\rm Re}\left[\frac{2 \lambda^{2+m}}{(1+m)(2+m)}
\,_2F_1\left(1, 1+\frac{m}{2};2+\frac{m}{2};\lambda^2\right)
\right]\, .
\eqne
We note the limiting forms of $F_m$ are
\eqnb
F_m(\lambda) \approx  \frac{2\lambda^{m} }{(1+m)} \left\lbrace \begin{array}{cc}
  \frac{\lambda^{2}}{2+m}&  \lambda \ll 1 \\
\frac{(-1) }{m} &  \lambda \gg 1 
\end{array}\right. \enspace ,
\eqne
where the long wavelength limit is only accurate for $m>1$.
\noindent 
This completes our description of the necessary equations to generate the full solution for an arbitrary order expansion. We conclude by writing out the first few terms:
\begin{align}
&K_0^m = \frac{3}{2} \frac{\lambda^{m+2}}{m+2}+\frac{3}{4}L_m +\epsilon\frac{3\pi {\rm i}}{4} \left\lbrace \begin{array}{cc}
   \frac{\lambda^{m+1}}{m+1} - \frac{\lambda^{m+3}}{m+3} &  |\lambda| < 1 \\
  \frac{1}{m+1} - \frac{1}{m+3} &  |\lambda| \geq 1 
\end{array}\right.  \nonumber\\
&K_1^m = \epsilon\left[ \frac{\lambda^{m+1}}{m+1} - K_0^{m+1}\right]~, \enspace \enspace
\label{K2}
K_2^m = K_0^{m+2} -  \frac{\lambda^{m+2}}{m+2}  \enspace .
\end{align}
Below we consider some specific integer examples which have special regular solutions.

\subsection{Specific solutions for integer values of $s$ }

We introduce a final recurrence relation, this time for $F_m$, valid for integer values of $m\geq 2$:
\begin{equation}
\label{recurrenceL}
(m+1)F_{m} = (m-1)F_{m-2} - 2 \frac{\lambda^m}{m}\enspace ,
\end{equation}
with 
$F_0 = -\ln|1-\lambda^2|$ and $F_1=\frac{1}{2}\ln\left|\frac{1+\lambda}{1-\lambda} \right| -\lambda$ .
With the above recurrence relations, the $K_n^m$ expression for different integer power-law exponents, are trivially found. As examples, we list below the expressions for the first few terms $m=0,1,2$, corresponding to $s=3, 4, 5$ respectively:

\begin{align}
&K_0^0 =  \frac{1}{2} \lambda^2 -\frac{1}{4}(\lambda+2)(\lambda-1)^2\ln\left|\frac{1+\lambda}{1-\lambda} \right|
 + \ln \left| 1+\lambda \right| + \epsilon\frac{\pi {\rm i}}{4} \left\lbrace \begin{array}{cc}
  \lambda(3- \lambda^2) &  |\lambda| < 1 \\
2 &  |\lambda| \geq 1 
\end{array}\right.  \nonumber \\
&K_1^0 =  \epsilon\left[\frac{1}{8} \lambda(5-3\lambda^2) +\frac{3}{16} \left(1-\lambda^2\right)^2\ln\left|\frac{1+\lambda}{1-\lambda} \right| \right] - \frac{3\pi {\rm i}}{16} \left\lbrace \begin{array}{cc}
  \lambda^2(2- \lambda^2) &  |\lambda| < 1 \\
1 &  |\lambda| \geq 1 
\end{array}\right.  \nonumber \\
&K_2^0 =   \left[-\frac{1}{10}\lambda^2\left( 4 - 3\lambda^2\right) - \frac{1}{20} \left(3 \lambda^5-5 \lambda^3+2\right)\ln\left|\frac{1+\lambda}{1-\lambda} \right| +\frac{1}{5}\ln\left|1+\lambda \right| \right] + \epsilon\frac{ \pi {\rm i}}{20}  \left\lbrace \begin{array}{cc}
  \lambda^3(5- 3 \lambda^2) &  |\lambda|< 1 \\
2 & |\lambda| \geq 1 
\end{array}\right.  \nonumber \\
&K_0^1 =  \frac{3}{8} \lambda(1+\lambda^2) -\frac{3}{16} \left(1-\lambda^2\right)^2\ln\left|\frac{1+\lambda}{1-\lambda} \right| + \epsilon\frac{3\pi {\rm i}}{16} \left\lbrace \begin{array}{cc}
  \lambda^2(2- \lambda^2) &  |\lambda| < 1 \\
1 &  |\lambda| \geq 1 
\end{array}\right.  \nonumber \\
&K_1^1 =  \epsilon \left[\frac{1}{10}\lambda^2\left( 4 - 3\lambda^2\right) + \frac{1}{20} \left(3 \lambda^5-5 \lambda^3+2\right)\ln\left|\frac{1+\lambda}{1-\lambda} \right| - \frac{1}{5}\ln\left|1+\lambda \right| \right]-  \frac{ \pi {\rm i}}{20}  \left\lbrace \begin{array}{cc}
  \lambda^3(5- 3 \lambda^2) &  |\lambda|< 1 \\
2 & |\lambda| \geq 1 
\end{array}\right.  \nonumber \\
&K_2^1= \frac{1}{24}\lambda\left(  6\lambda^4-7\lambda^2 +3 \right) - \frac{1}{16} \left(2 \lambda^6-3 \lambda^4 +1\right)\ln\left|\frac{1+\lambda}{1-\lambda} \right|  +  \epsilon\frac{ \pi {\rm i}}{16} \left\lbrace \begin{array}{cc}
  \lambda^4(3-2 \lambda^2) &  |\lambda| < 1 \\
1 &  |\lambda| \geq 1 
\end{array}\right.  \nonumber   
\\
&K_0^2 = \frac{1}{10}\lambda^2\left( 1+ 3\lambda^2\right) - \frac{1}{20} \left(3 \lambda^5-5 \lambda^3+2\right)\ln\left|\frac{1+\lambda}{1-\lambda} \right| + \frac{1}{5}\ln\left|1+\lambda \right| + \epsilon \frac{ \pi {\rm i}}{20}  \left\lbrace \begin{array}{cc}
  \lambda^3(5- 3 \lambda^2) &  |\lambda|< 1 \\
2 & |\lambda| \geq 1 
\end{array}\right.  \nonumber \\
&K_1^2 = \epsilon \left[-\frac{1}{24}\lambda\left(  6\lambda^4-7\lambda^2 +3 \right) + \frac{1}{16} \left(2 \lambda^6-3 \lambda^4 +1\right)\ln\left|\frac{1+\lambda}{1-\lambda} \right| \right] -\frac{ \pi {\rm i}}{16} \left\lbrace \begin{array}{cc}
  \lambda^4(3-2 \lambda^2) &  |\lambda| < 1 \\
1 &  |\lambda| \geq 1 
\end{array}\right.  \nonumber  \\
&K_2^2 =  \frac{1}{140} \left[2\lambda^2\left( 3 - 16\lambda^2+15 \lambda^4\right) -3 \left(5 \lambda^7-7 \lambda^5+2\right)\ln\left|\frac{1+\lambda}{1-\lambda} \right| +12\ln\left|1+\lambda \right| \right] +\epsilon  \frac{ 3 \pi {\rm i}}{140}  \left\lbrace \begin{array}{cc}
  \lambda^5(7- 5 \lambda^2) &  |\lambda|< 1 \\
2 & |\lambda| \geq 1 
\end{array}\right.  \nonumber
\end{align}

\subsection{Moments of the distribution function}

To help visualise the above results, normalised by the unhelpful $\phi_\ell$ terms, it is convenient to introduce more familiar fluid concepts. Using the ortho-normality of the Legendre polynomials, one can easily recover the cosmic-ray density, current and pressure tensor, being respectively:
\begin{align*}
n_{\rm cr} &= \int d^3p~ f =  4 \pi\int dp~p^2 f_0 \\
j_{\rm cr} &= q\int d^3p~ v_x f  = \frac{4\pi}{3} q\int dp~p^2 v f_1 \\
T^{\mu\nu}_{\rm cr} &= \int d^3p^\mu v^{\nu} f =   \frac{4\pi}{3} \int dp~p^3 v \left[f_0\delta^{\mu\nu} 
+ \frac{f_2}{5}{\rm diag}\left(2 ,-1,-1 \right)\right]  
\end{align*}
From the latter, we identify the isotropic and anisotropic pressures
\begin{align*}
&P_{\rm cr} =   \frac{4\pi}{3} \int dp~p^3 v f_0 ~, \enspace \enspace
\Delta P_{\rm cr}\equiv T_{\rm cr}^{11}-T_{\rm cr}^{22} =   \frac{4\pi}{5} \int dp~p^3 v f_2 \enspace .
\end{align*}
For power-law distributions as we have used in the previous section, we can re-express the first few $\phi_n$ in terms of these fluid parameters:

 \begin{equation*}
  \frac{4 \pi}{3n_{\rm cr}}\phi_0 = \frac{1}{3} \alpha = \frac{ P_{cr}}{n_{\rm cr}c}\kappa\enspace \enspace,\enspace
 \frac{4 \pi}{3n_{\rm cr}} \phi_1  = \frac{v_{\rm cr}}{c}\alpha\enspace\enspace,\enspace
  \frac{4 \pi}{3n_{\rm cr}} \phi_2  = \frac{5}{9}\frac{\Delta P_{cr}}{P_{cr}}\alpha
 \end{equation*}

 where 
  \begin{equation*}
 \alpha=\left\lbrace 
 \begin{array}{cc}
 \left[ \ln\left(\frac{p_2}{p_1}\right)\right]^{-1}  & s=3 \\  (s-3)\left[p_1^{3-s} - p_2^{3-s}\right]^{-1} & s\neq3
 \end{array}\right. ~~\mbox{and}~~~~~
  \kappa=\left\lbrace 
 \begin{array}{cc}
 \left[ \ln\left(\frac{p_2}{p_1}\right)\right]^{-1}  & s=4 \\  (s-4)\left[p_1^{4-s} - p_2^{4-s}\right] ^{-1}& s\neq4
 \end{array}\right.
 \end{equation*}
are the relevant momentum integrals.

\begin{figure}
	\begin{center}
		\includegraphics[width=0.9\textwidth]{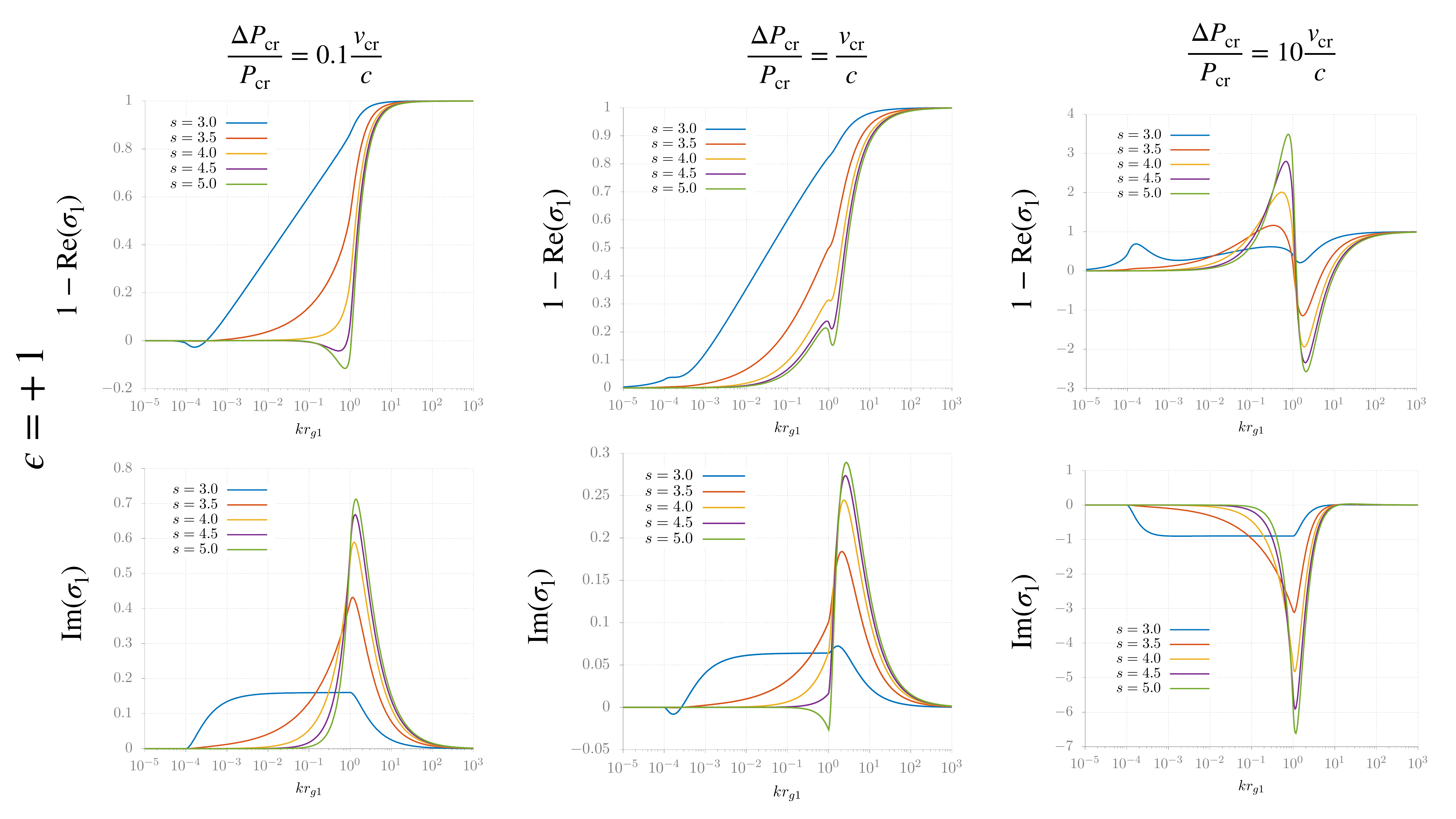}
		\includegraphics[width=0.9\textwidth]{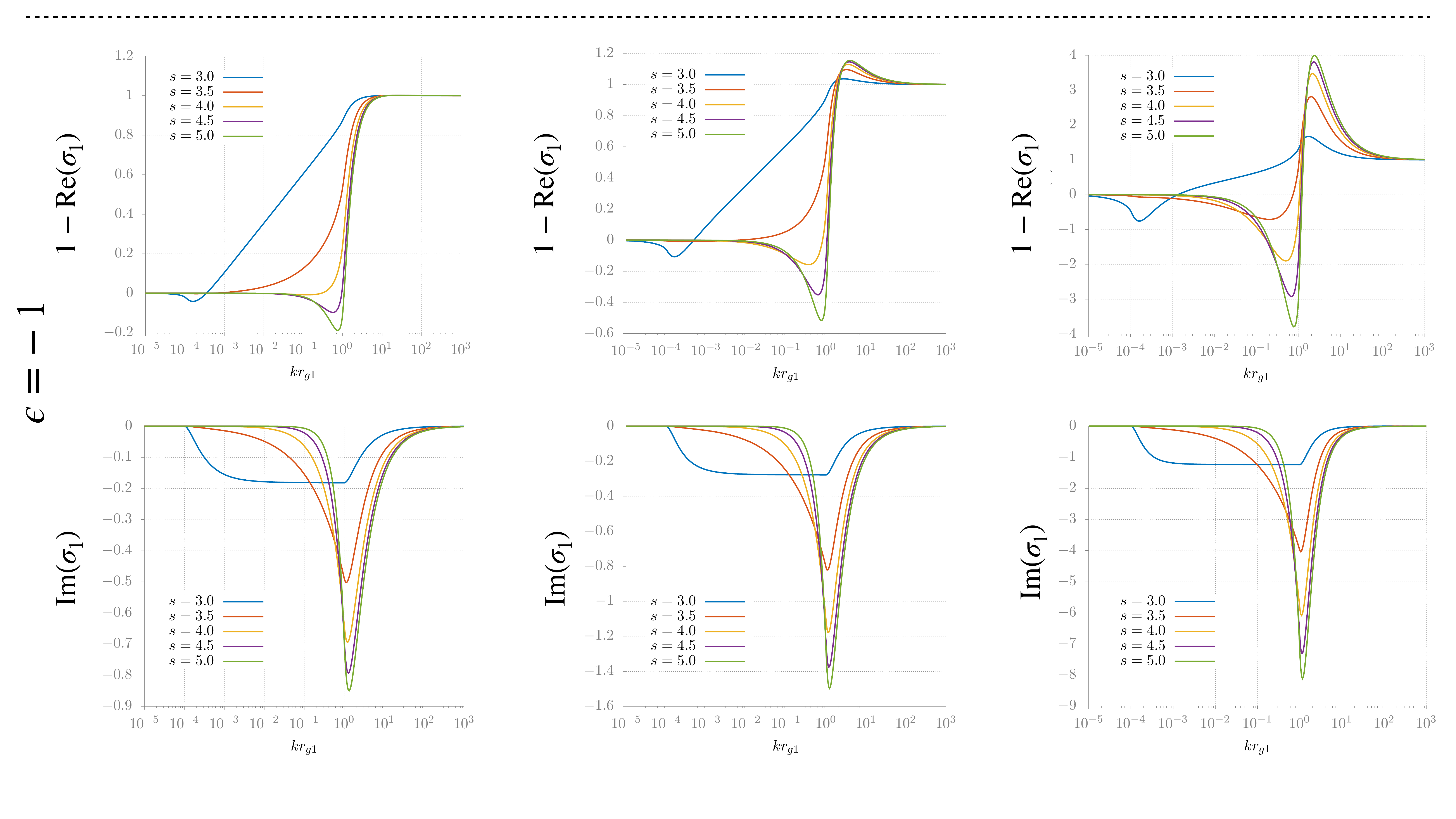}
	\end{center}
	\caption{
		$\sigma_1$ plotted for different power-law indices $s$. In all cases  $p_2/p_1 = 10^4$. 
		For the real part, $1-{\rm Re}\sigma$ is plotted, to emphasise how it approaches zero as $k\rightarrow 0$, as it must. 
	}
	\label{fig:A1}
\end{figure}

 \subsection{Susceptibilities - example to second order in $f_\ell$ and limiting cases}
 
Using these expressions, we provide the required scalar susceptibilities, i.e. the response of the CR current to linear magnetic perturbations. The relative magnitudes of the different moments will depend on the physical scenario under investigation. For example, in the diffusion approximation, it is generally found that $\phi_{\ell+1} \approx (v_{\rm cr}/c)\phi_{\ell}$ \cite[e.g.][]{JokipiiWilliams}.

As we have only considered the first three fluid moments, we consider only the terms in the Legendre expansion up to $f_2$, in which case
\begin{align}
&\sigma_1 =\alpha \left(\lambda_1 p_1\right)^{3-s}\left[ K_0^{s-3} + \frac{5}{3} \frac{c}{v_{\rm cr}}\frac{\Delta P_{cr}}{P_{cr}} K_1^{s-3}\right]_{\lambda_2}^{\lambda_1} \nonumber \\
&{\rm  and}  \label{Eq:sigma2} \\
&\sigma_2  =  \frac{\alpha}{3}  \left(\lambda_1 p_1\right)^{3-s}\left[  
s K_0^{s-3} - \lambda^{s-2} I_0  +  3 \frac{v_{\rm cr}}{c}  \left( (s+1) K_1^{s-3}-\lambda^{s-2} I_1\right) 
+ \frac{5}{6} \frac{\Delta P_{cr}}{P_{cr}} 
 \left(  \lambda^{s-2} \left(I_0   -3I_2 \right)- s  K_0^{s-3} +3(s+2) K_2^{s-3} \right) 
 \right]_{\lambda_2}^{\lambda_1} \nonumber
\end{align}
where expressions for $K_n^m$ are found in equations (\ref{K2}) above.  

The real and imaginary parts of $\sigma_1$ are plotted in Figure \ref{fig:A1}, where we have taken $\Delta P_{\rm cr} > 0$ in all cases. A crucial test of the solution is that the long wavelength limit ($kr_{g2} \ll 1$), the cosmic rays must follow the magnetic fields, i.e. $j_\bot/j_\| = B_\bot/B_\|$, or equvalently, $\sigma \rightarrow 1$. We observe that this is indeed the case.
  $\sigma_2$ is dominated by the leading two terms in equation \ref{Eq:sigma2}, and plotting for different parameters is un-necessary, although graphically $\sigma_2$ look very similar to the left column in Figure \ref{fig:A1}. This is also to be expected, since it is this term which stabilises the resonant streaming instability when $\omega/v_{\rm cr} k$ approaches unity,  which in turn determines the neutral streaming speed of the cosmic rays. Strong anisotropy of course changes this \cite[see for example][]{Zweibel20}.

 \subsection{Long-wavelength approximations}
 
 For $\lambda \gg 1$  one can show for non-integer $m$
\begin{align*}
&K_0^m \approx  \frac{\lambda^m}{m}+\frac{1}{5}  \frac{\lambda^{m-2}}{m-2}
   +\epsilon \frac{3\pi {\rm i}}{4}\left[ \frac{1}{1+m}- \frac{1}{3+m}\right]\\
& K_1^m \approx -\frac{\epsilon}{5}  \frac{\lambda^{m-1}}{m-1}
   - \frac{3\pi {\rm i}}{4}\left[ \frac{1}{2+m}- \frac{1}{4+m}\right]\\
& K_2^m  \approx  \frac{1}{5}  \frac{\lambda^{m}}{m}
   +\epsilon \frac{3\pi {\rm i}}{4}\left[ \frac{1}{3+m}- \frac{1}{5+m}\right]\\
 &I_0 \approx \lambda^{-1}, ~~~~~ I_1 ~~\approx~~ - \frac{\epsilon}{5}  \lambda^{-2}, ~~~~~ I_2 ~~\approx ~~\frac{1}{5}  \lambda^{-1}
\end{align*}

Integer cases can be found from the final expressions using L'H\^opital's rule. 
 
\subsubsection{Non-resonant long-wavelength}
For $\lambda_{1} \gg \lambda_2\gg 1 $ the imaginary parts of the integrals vanish, while to order $k^2$ 
\begin{equation}
\label{sig1-lw}
  1-\sigma_1 =k r_{g,1}\left[ \epsilon\frac{(s-3)}{3(s-4)} \frac{c}{v_{\rm cr}}\frac{\Delta P_{cr}}{P_{cr}}
 \left( \frac{1-\left(\frac{p_1}{p_2}\right)^{s-4}}{1-\left(\frac{p_1}{p_2}\right)^{s-3}}  \right)
  -\frac{(s-3)}{5(s-5) } \left(\frac{1-\left(\frac{p_1}{p_2}\right)^{s-5}}{1-\left(\frac{p_1}{p_2}\right)^{s-3}} \right)
    k r_{g,1} \right]
 \end{equation}
 and \begin{equation}
  1-\sigma_2 = k r_{g,1}\left[ \epsilon \frac{s-3}{s-4}\frac{v_{\rm cr}}{c} 
   \left(\frac{1-\left(\frac{p_1}{p_2}\right)^{s-4}}{1-\left(\frac{p_1}{p_2}\right)^{s-3}} \right)
  -\frac{s}{15}\frac{s-3}{s-5}\left(1-\frac{5}{6}\frac{\Delta P_{cr}}{P_{cr}} \right)
     \left(\frac{1-\left(\frac{p_1}{p_2}\right)^{s-5}}{1-\left(\frac{p_1}{p_2}\right)^{s-3}} \right)    k r_{g,1} \right]
 \end{equation}
 From these, the standard firehose instability result follows.

\subsubsection{Resonance}
For $\lambda_{1} > 1 > \lambda_2 $  resonance is possible. We simplify the picture by taking the limit of $\lambda_2\rightarrow 0$ and $\lambda_{1} \gg 1 $,  finding
\begin{equation}
\label{sig1-res}
  1-\sigma_1 =k r_{g,1}\left[ \epsilon\frac{(s-3)}{3(s-4)} \frac{c}{v_{\rm cr}}\frac{\Delta P_{cr}}{P_{cr}}
  -\frac{(s-3)}{5(s-5) }  k r_{g,1} \right] - \frac{3 \pi {\rm i}}{2} \left(k r_{g,1}\right)^{s-3}
  \left[ \frac{\epsilon(s-3)}{s(s-2)} -\frac{5(s-3)}{3(s^2-1)} \frac{c}{v_{\rm cr}}\frac{\Delta P_{cr}}{P_{cr}}\right]
 \end{equation}
  and \begin{equation}
  \label{sig2-res}
  1-\sigma_2 = k r_{g,1}\left[ \epsilon \frac{s-3}{s-4}\frac{v_{\rm cr}}{c} 
  -\frac{s}{15}\frac{s-3}{s-5}\left(1-\frac{5}{6}\frac{\Delta P_{cr}}{P_{cr}} \right)   k r_{g,1} \right]
  - \frac{ \pi {\rm i}}{2} \left(k r_{g,1}\right)^{s-3}
  \left[ \frac{\epsilon(s-3)}{(s-2)} -\frac{3}{(s-1)} \frac{v_{\rm cr}}{c}+
  \frac{5}{6}\frac{\Delta P_{cr}}{P_{cr}}  \left( \frac{3}{s}- \frac{\epsilon}{s-2}\right)\right]
 \end{equation}
Since the pressure diverges for $s<4$, this result only applies to cases in which $s>4$. For harder spectra, one can combine with the previous results for $\lambda_2\ll1$, in which case the leading order terms effectively cancel. i.e. the Firehose instability operates only in the long-wavelength, non-resonant limit.

\bsp	
\label{lastpage}
\end{document}